\documentclass[%
reprint,
superscriptaddress,
 amsmath,amssymb,
 aps,
 prl,
]{revtex4-1}

\usepackage{graphicx}
\usepackage{hyperref}
\usepackage[export]{adjustbox}
\usepackage{tikz}
\usepackage{braket}
\usepackage{color, ulem}
\usepackage{textcomp,gensymb}
\usepackage{dsfont}

\newcommand{\makeauthor}[3]{%
  \newcommand{#1}[1]{%
    {\sffamily\color{#3}{\bfseries #2:} ##1}%
  }%
}

\makeauthor{\lk}{Lennart}{blue}

\newcommand{\bvec}[1]{\boldsymbol #1}
\newcommand{\Ucrit}{U_\mathrm{c}}

\begin{document}

\title{Importance of long-ranged electron-electron interactions for the magnetic phase diagram of twisted bilayer graphene}

\author{Lennart Klebl}
\affiliation{Institute for Theory of Statistical Physics, RWTH Aachen University, and JARA Fundamentals of Future Information Technology, 52062 Aachen, Germany\\}
\author{Zachary A. H. Goodwin}
\affiliation{Departments of Materials and Physics and the Thomas Young Centre for Theory and Simulation of Materials, Imperial College London, South Kensington Campus, London SW7 2AZ, UK\\}
\author{Arash A. Mostofi}
\affiliation{Departments of Materials and Physics and the Thomas Young Centre for Theory and Simulation of Materials, Imperial College London, South Kensington Campus, London SW7 2AZ, UK\\}
\author{Dante M. Kennes}
\affiliation{Institute for Theory of Statistical Physics, RWTH Aachen University, and JARA Fundamentals of Future Information Technology, 52062 Aachen, Germany\\}
\affiliation{Max Planck Institute for the Structure and Dynamics of Matter,
Center for Free Electron Laser Science, 22761 Hamburg, Germany\\}
\author{Johannes Lischner}
\affiliation{Departments of Materials and Physics and the Thomas Young Centre for Theory and Simulation of Materials, Imperial College London, South Kensington Campus, London SW7 2AZ, UK\\}

\date{\today}

\begin{abstract}
Electron-electron interactions are intrinsically long-ranged, but many models of strongly interacting electrons only take short-ranged interactions into account. Here, we present results of atomistic calculations including both long-ranged and short-ranged electron-electron interactions for the magnetic phase diagram of twisted bilayer graphene and demonstrate that qualitatively different results are obtained when long-ranged interactions are neglected. In particular, we use Hartree theory augmented with Hubbard interactions and calculate the interacting spin susceptibility at a range of doping levels and twist angles near the first magic angle to identify the dominant magnetic instabilities. At the magic angle, mostly anti-ferromagnetic order is found, while ferromagnetism dominates at other twist angles. Moreover, long-ranged interactions significantly increase the twist angle window in which strong correlation phenomena can be expected. These findings are in good agreement with available experimental data.
\end{abstract}

\maketitle

\section{Introduction}

Since the discovery of superconductivity in proximity to correlated insulator states at half (electron or hole) filling of the flat bands~\cite{NAT_I,NAT_S}, there has been great interest in the electronic properties of magic-angle twisted bilayer graphene (tBLG)~\cite{TT}. Additional experiments~\cite{TSTBLG,SOM,cao2020nematicity,Zondiner2020,Vafek2020,Wong2020,NAT_SS} discovered correlated insulator phases and superconductivity at other doping levels of the flat bands and revealed a wide range of interesting phenomena~\cite{Balents2020,moiresim} including strange metal behaviour~\cite{SMTBLG,Polshyn2019}, ferromagnetic order~\cite{EFM,Serlin2020}, superconductivity without correlated insulators~\cite{Saito2020,Arora2020,Stepanov2020}, Chern insulators~\cite{Chern_Das,Chern_Wu,Chern_Nuckolls}, and nematic order \cite{NAT_MEI,NAT_CO,IEC,cao2020nematicity}.

These findings demonstrate the importance of electron-electron interactions for understanding the electronic properties of tBLG~\cite{Balents2020,moiresim}. The quintessential model for strongly interacting electrons is the Hubbard model, in which electrons only interact when they are on the same ``site" (typically assumed to be an atom). In tBLG near the magic angle, the moir\'e pattern results in the emergence of eight flat bands (including a factor of two from spin degeneracy) near the Fermi energy which are separated from all other bands by energy gaps~\cite{GBWT,MBTBLG,LDE,NSCS,Cantele2020,STBBG,SETLA,CMLD,KDP}. Starting from atomistic tight-binding approaches, Hubbard models for tBLG can be obtained by constructing Wannier functions of the flat bands~\cite{MLWO,SMLWF,PHD_1} (note that this is not possible when a continuum model starting point is used; in that case additional bands must be included in the Wannierization procedure~\cite{FTBM,Carr2019wannier}). The properties of such models have been studied using mean-field theory, the functional renormalization group~\cite{SCDID,LK_DMK_CH} and exact diagonalization~\cite{NAT_SS} resulting in many important insights into the origin of superconductivity and correlated insulator states. Instead of using flat-band Wannier functions which are extended over the whole moir\'e unit cell as a basis, it is also possible to construct atomistic Hubbard models using a basis of carbon $p_z$ orbitals~\cite{ECM,LK_CH,Ramires2019}. 

Importantly, Hubbard models only capture short-ranged electron-electron interactions~\footnote{Note that Hubbard models using a basis of flat-band Wannier functions account for some long-ranged interactions because of the large size of the Wannier orbitals~\cite{MLWO,SMLWF,PHD_1}}. It is well known, however, that long-ranged interactions play an important role in tBLG. Using Hartree theory, several groups~\cite{EE,Cea2019,Rademaker2019,PHD_4,Bascones2020} demonstrated that long-ranged interactions result in significant changes of the electronic structure which depend sensitively on doping and twist angle. In particular, Hartree interactions result in a flattening of the doped bands (in addition to the band flattening induced by twisting)~\cite{GBWT,MBTBLG,LDE,NSCS}. This interaction-induced band flattening explains the Fermi level pinning that was observed in several recent scanning tunnelling spectroscopy measurements~\cite{NAT_SS,NAT_MEI}.

Therefore, accurate models of tBLG should capture both short-ranged and long-ranged electron-electron interactions. To achieve this, several groups used Hartree-Fock calculations based on a continuum model of the electronic structure~\cite{SCHFC,Bultinck2020,Liu2019,Zhang2020,Cea2020}. While these calculations have yielded many useful insights, they do not capture atomic-scale interactions (such as onsite interactions within carbon $p_z$-orbitals) and often only include a few bands near the Fermi level with the effect of all other bands being described by an effective dielectric constant. Few groups have attempted to capture the interplay of long-ranged and short-ranged interactions using atomistic calculations: Gonz\'alez and Stauber~\cite{Stauber2020} studied the properties of tBLG in different dielectric environments using atomistic Hartree-Fock theory, and Sboychakov \textit{et al.}~\cite{Sboychakov2019,Sboychakov2020} developed an atomistic Hubbard model with electron-electron interactions beyond the atomistic Hubbard interactions. These studies investigated the properties of tBLG at a single twist angle, and therefore, did not study in detail the doping and twist-angle dependence of the interplay of long-ranged and short-range interactions.

In this paper, we calculate the magnetic phase diagram of tBLG as function of twist angle and doping using an atomistic Hartree theory with additional Hubbard interactions. Specifically, we calculate the interacting spin susceptibility and determine the critical value of the Hubbard $U$ parameter that is required to induce a magnetic instability. Our calculations predict magnetic instabilities over a relatively large twist angle window ranging from $0.96\degree$ to $1.16\degree$. Near the magic angle, the magnetic ordering is mostly anti-ferromagnetic, while at other twist angles ferromagnetism dominates. When Hartree interactions are neglected, a qualitatively different phase diagram with a much smaller critical twist angle window is found. Finally, we compare our findings with available experimental data and overall find good agreement.

\begin{figure*}[ht]
\centering
\begin{minipage}[b]{0.32\textwidth}
\centering 
\includegraphics[width=\textwidth]{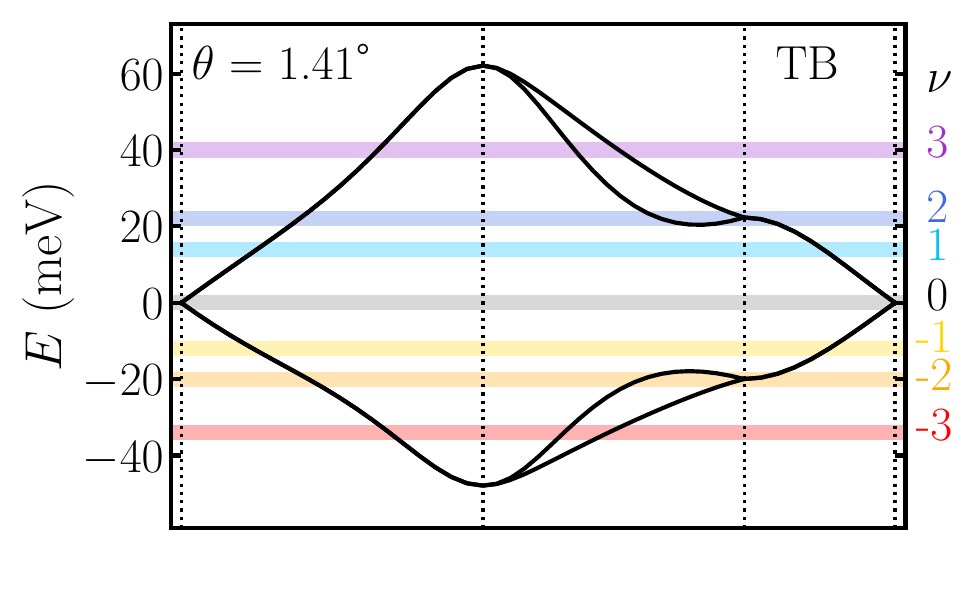}
\end{minipage}
\begin{minipage}[b]{0.32\textwidth}
\centering 
\includegraphics[width=\textwidth]{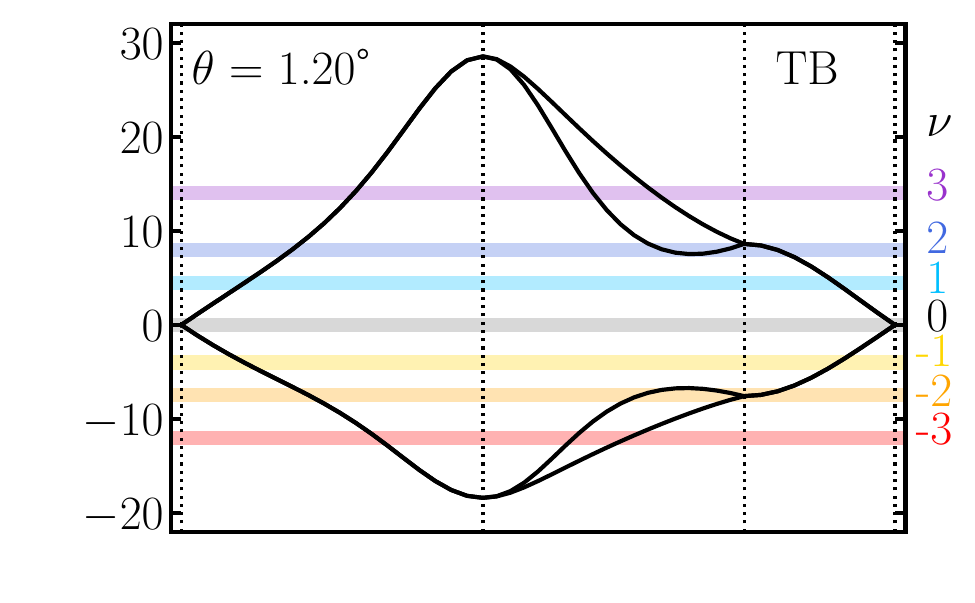}
\end{minipage}
\begin{minipage}[b]{0.32\textwidth}   
\centering
\includegraphics[width=\textwidth]{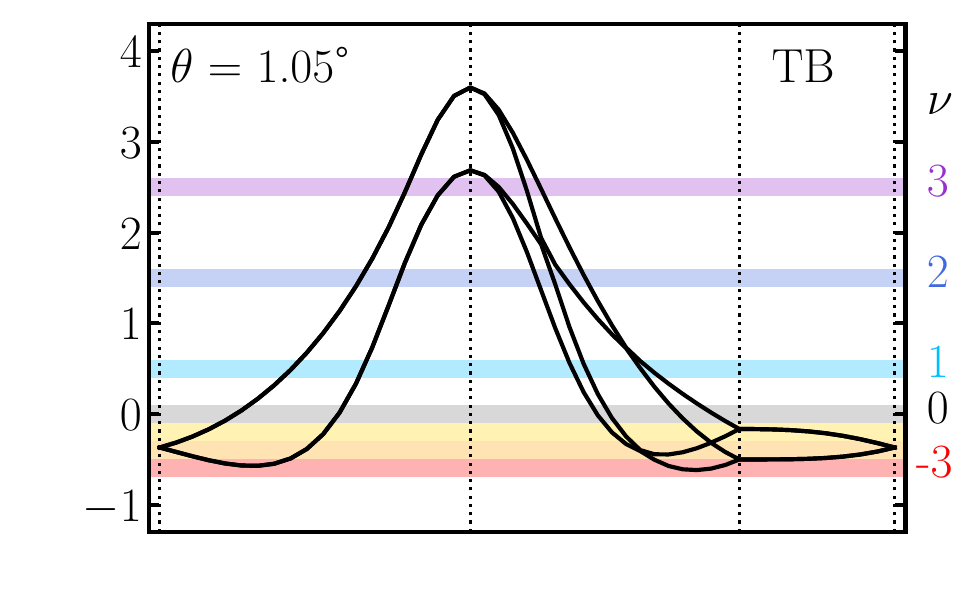}
\end{minipage}
\centering
\begin{minipage}[b]{0.32\textwidth}
\centering 
\includegraphics[width=\textwidth]{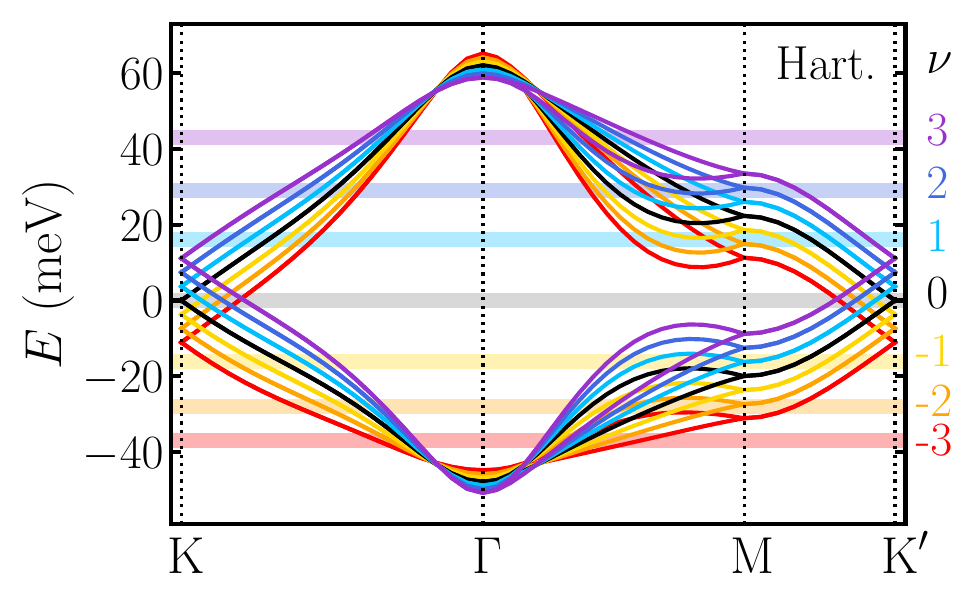}
\end{minipage}
\begin{minipage}[b]{0.32\textwidth}
\centering 
\includegraphics[width=\textwidth]{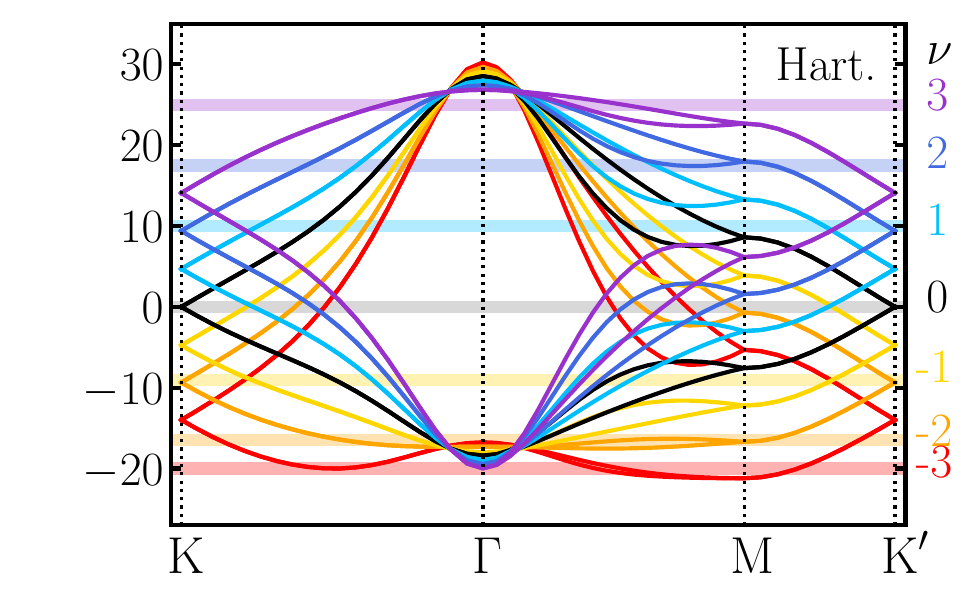}
\end{minipage}
\begin{minipage}[b]{0.32\textwidth}   
\centering 
\includegraphics[width=\textwidth]{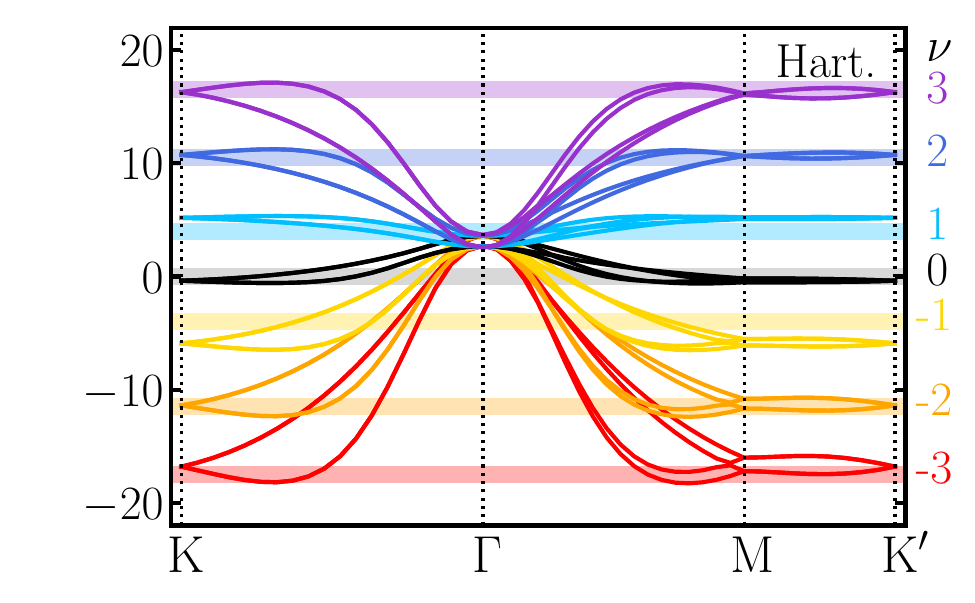}
\end{minipage}
\caption{Band structures of tBLG at twist angles of $1.41\degree$, $1.20\degree$ and $1.05\degree$ for integer fillings $\nu$ of the flat bands from tight-binding (denoted TB, see upper panels) and Hartree theory (denoted Hart., see lower panels). Fermi levels are indicated by horizontal lines. In contrast to Hartree theory, the tight-binding band structure does not depend on $\nu$. Note that the energy scale on the $y$-axis is different in each panel. The zero of energy for each plot is taken to be the Dirac point energy from tight-binding.} 
\label{fig:BS_H}
\end{figure*}

\section{Methods}
We study commensurate unit cells of tBLG with $D_3$ symmetry~\cite{NSCS,LDE}. The atomic positions are relaxed using classical force-fields~\cite{Cantele2020,STBBG,SETLA,CMLD,KDP,PHD_5}. For this, we use a combination of the AIREBO-Morse~\cite{AIREBO} and Kolmogorov–Crespi~\cite{KC} potentials as implemented in LAMMPS~\cite{LAMMPS}.

To investigate magnetic ordering tendencies of tBLG including the effect of long-ranged interactions, we calculate the interacting static spin susceptibility in the normal state using a Hartree theory plus $U$ (Hartree$+U$) approach. In this approach, Hubbard interactions within the carbon $p_z$-orbitals are captured by adding a Hubbard contribution $U\sum_i n_{i\uparrow} n_{i\downarrow}$ (with $U$ denoting the Hubbard parameter and $n_{i\uparrow}$($n_{i\downarrow}$) denoting the occupancy of the up(down)-spin $p_z$-orbital on carbon atom $i$) to the Hartree theory total energy. This approach assumes that the spatial range of the exchange interaction is strongly reduced as a result of electronic screening induced by the flat bands~\cite{PHD_2,CCRPA,Cea2021}. Moreover, it has been shown that models with short-ranged Hubbard-type exchange interactions accurately describe the magnetic phase diagram of graphene and bilayer graphene bilayer~\cite{Scherer2012,Lang2012,Honerkamp2008}, which can be viewed as ``parent" systems whose ordering tendencies are inherited by the twisted bilayer graphene~\cite{LK_CH}.

The Hartree Hamiltonian is given by 
\begin{equation}
    \hat{\mathcal H}_H = \sum_{ij}\,t(\bvec\tau_i-\bvec\tau_j)\,
  \hat{c}^\dagger_i \hat{c}^{\phantom{\dagger}}_j + \sum_{i}V(\bvec\tau_i)\,\hat{c}^{\dagger}_{i}\hat{c}_{i}^{\phantom{\dagger}},
\label{eq:HH}
\end{equation}
where $\boldsymbol{\tau}_i$ is the position vector of carbon atom $i$ and the corresponding annihilation (creation) operators are denoted by $\hat{c}^{(\dagger)}_{i}$. The hopping parameters $t(\bvec r)$ are determined using the Slater-Koster rules~\cite{NSCS,LDE,SK}. For this, we use the parameterization from Refs.~\onlinecite{NSCS} and~\onlinecite{Moon2012}
\begin{equation}
t(\bvec{r}) = t_{\sigma}e^{q_{\sigma}(1 - |\bvec{r}|/d)}\cos^{2}\varphi + t_{\pi}e^{q_{\pi}(1 - |\bvec{r}|/a)}\sin^{2}\varphi,
\end{equation}
where $t_{\sigma} = 0.48\,\mathrm{eV}$ and $t_{\pi} = -2.7\,\mathrm{eV}$ are, respectively, the sigma and pi hopping between carbon $p_z$ orbitals, and $d = 3.3\,\text{\AA}$ and $a = 1.4\,\text{\AA}$ denote the interlayer separation and carbon-carbon bond length, respectively. Also, $q_{\sigma} = d/(0.184a)$ and $q_{\pi} = 1/0.184$ are decay parameters and $\varphi$ is the inclination angle of the orbitals.

The second term in Eq.~\eqref{eq:HH} describes long-ranged Hartree interactions with $V(\bvec\tau_i)$ denoting the Hartree potential at position $\bvec\tau_i$. The Hartree potential is given by 
\begin{equation}
V(\bvec\tau_i) = \sum_j (n_j - \overline{n}) W_{ij},
\end{equation}
where $n_j$ denotes the occupancy of the $p_z$ orbital on atom $j$ and $\overline{n}$ is the average occupancy~\cite{Rademaker2019}. Also, $W_{ij}$ denotes the screened Coulomb interaction between electrons at $\bvec\tau_i$ and $\bvec\tau_j$~\cite{PHD_4}. In principle, $V(\bvec\tau_i)$ must be determined self-consistently, but it has been shown~\cite{EE,Cea2019,PHD_4} that the resulting potential is accurately described by
\begin{equation}
V(\bvec\tau_i) = (\nu - \nu_0) V_0\sum_{j=1,2,3}\cos(\bvec{b}_j\cdot\boldsymbol{\tau}_i),
\label{eq:fit}
\end{equation}
where $\nu$ is the number of added electrons per moir\'e unit cell (relative to charge neutrality) and $\bvec b_j$ denote the three shortest reciprocal lattice vectors of the moir\'e unit cell. We use $V_0$ = $5\,\mathrm{meV}$  and $\nu_0=0$~\cite{PHD_4}. These parameters include internal screening from tBLG~\cite{CCRPA,PHD_2}. Additional screening from the substrate or metallic gates~\cite{PHD_3,PHD_4,Bascones2020} results in a further reduction of $V_0$. Note that Eq.~\eqref{eq:fit} assumes that the AA regions reside in the corners of the rhombus-shaped moir\'e unit cell.

Within the Hartree$+U$ approach, the frequency- and wavevector-dependent interacting spin susceptibility $\chi_{ij}(\bvec{q},q_0)$ (with $i$ and $j$ denoting carbon $p_z$-orbitals and $\bvec{q}$ and $q_0$ being a wavevector and frequency, respectively) is given by~\cite{LK_CH,Lischner2015} 
\begin{align}
    \hat{\chi}(\bvec q,q_0) &{}= \hat{\chi}^{(0)}(\bvec q,q_0)\,\big[\mathds{1}+U\hat{\chi}^{(0)}(\bvec q,q_0)\big]^{-1}.
\end{align}
Here, $\hat{\chi}^{0}$ denotes the non-interacting spin-response function
\begin{align}
    \chi_{ij}^{(0)}(\bvec q,q_0) &{}= \frac1{N_{\bvec k}\beta}\,\sum_{\bvec k,k_0}
        G_{ij}(\bvec k,k_0)G_{ji}(\bvec k+\bvec q,k_0+q_0), 
\end{align}
where $\hat G(\bvec k,k_0) = (ik_0 - \hat{\mathcal{H}}_H(\bvec k) + \mu)^{-1}$ is the Matsubara Green's function of the Hartree Hamiltonian for states with crystal momentum $\bvec{k}$. Also, $\mu$ denotes the chemical potential, $N_{\bvec{k}}$ is the number of momentum points used to sample the first Brillouin zone ($N_{\bvec k}=24$) and $\beta=1/(k_\mathrm{B} T)$ (with $k_\mathrm{B}$ and $T$ denoting the Boltzmann constant and temperature, respectively). The Matsubara ($k_0$) summation is carried out numerically using an appropriately chosen grid with $N_\omega=500$ frequencies %
\footnote{The frequency grid is chosen to both be linearly spaced for $n \ll N_\omega$ ($n$ denotes the positive Matsubara frequency index ranging from zero to $N_\omega-1$) and increase its spacing quadratically for higher frequencies. This is achieved by $\omega_n\propto\tan[z_n\,\pi/2\,(2n+1)/(2N_\omega-1)]$ with $z_n=1-\epsilon_z [n/(N_\omega-1)]^{\alpha_z}$ controlling the strength of the divergence. In our case, we set $\epsilon_z = 10^{-7}$ and $\alpha_z=5$. The proportionality factor is chosen such that for small $n$, the original Matsubara frequencies are reproduced. The corresponding weights are determined by the derivative of the above formula with respect to $n$.}
-- this reduces the computational effort compared to the analytical evaluation~\cite{LK_CH}. For comparison, we also calculate the interacting spin susceptibility without long-ranged interactions, i.e., setting $V(\bvec\tau_i)=0$. 

We focus on low-temperature static instabilities that maintain the translational symmetry of the moir\'e lattice and therefore use $\beta = 10^4\,\mathrm{eV}^{-1}$ and $\bvec q = q_0 = 0$.
Magnetic instabilities occur when an eigenvalue of $\chi_{ij}$ diverges. The critical interaction strength that is required to induce the ordering is given by $\Ucrit=-1/\lambda_0$, where $\lambda_0$ denotes the largest eigenvalue of $\chi^{(0)}_{ij}$. This is a generalization of the well-known Stoner criterion of ferromagnetism~\cite{LinHirschFM}. The corresponding eigenvector $v_i$ of $\chi_{ij}$ characterizes the spatial structure of the resulting magnetic order.

\section{Results}

Figure~\ref{fig:BS_H} shows the band structures from Hartree theory at three twist angles near the magic angle ($\theta=1.41\degree$, $1.20\degree$ and $1.05\degree$) at various doping levels. For comparison, we also show the corresponding tight-binding results. For the two larger twist angles, both the Hartree and tight-binding band structures exhibit Dirac cones at the K and K$^\prime$ points. While the non-interacting tight-binding band structure does not depend on the doping level, long-ranged electron-electron interactions captured by Hartree theory give rise to a significant doping-dependent distortion of the band structure~\cite{EE,Cea2019,Rademaker2019,PHD_4,Bascones2020}. In particular, Hartree interactions result in a flattening of the doped bands. For example, at $\theta=1.20\degree$ and $\nu=3$ the two higher-energy bands are much flatter than the corresponding tight-binding bands. 

The magic angle (defined as the twist angle with the smallest width of the flat band manifold from tight-binding) is found to be $1.05\degree$. At this twist angle, the tight-binding band structure differs qualitatively from the result at larger (and smaller) twist angles. In particular, the lower-energy bands are inverted and have a similar shape to the higher-energy bands. Including long-ranged interactions again results in drastic changes to the band structure with Hartree theory predicting an increase of the overall flat band width when the system is doped. Also, the overall shape of the flat band manifold is flipped when comparing hole-doped and electron-doped systems. 

The strong band deformations which are observed in the doped tBLG can be understood from an analysis of the electron wavefunctions in real space~\cite{Rademaker2019,EE}. At the centre of the Brillouin zone, near $\Gamma$, the wavefunctions are localized in the AB and BA regions of the moir\'e unit cell. In contrast, states near M and K/K' are localized in the AA regions of the moir\'e unit cell. When tBLG is doped, states near K/K' are first populated by electrons (or holes) resulting in an inhomogeneous charge density which gives rise to a strong Hartree potential. The Hartree potential, in turn, interacts strongly with states which are localized in the AA regions resulting in an energy shift of states near M and K/K', while states near $\Gamma$ are less strongly affected. There are small relative distortions between the M and K point because they are localised in a similar manner in the AA regions. For more detailed discussions of the Hartree-theory band structures, we refer the interested reader to  Refs.~\citenum{EE,Cea2019,PHD_4,Rademaker2019,Bascones2020}.


\begin{figure}[th]
\centering
\includegraphics[width=0.98\columnwidth]{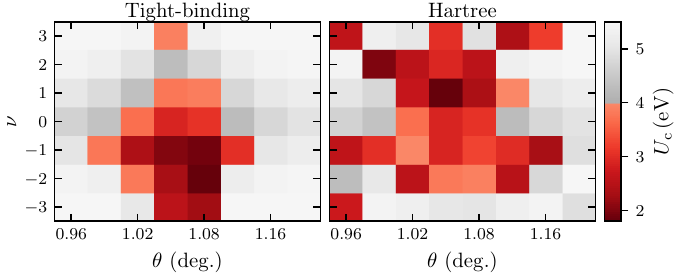}
\caption{Critical Hubbard interaction strength $\Ucrit$ required for the onset of magnetic instabilities as a function of flat band filling $\nu$ and twist angle $\theta$. Left panel: without Hartree interactions (tight-binding). Right panel: with Hartree interactions.}
\label{fig:CU}
\end{figure}

\begin{figure}[th]
\centering
\includegraphics[width=0.9\columnwidth]{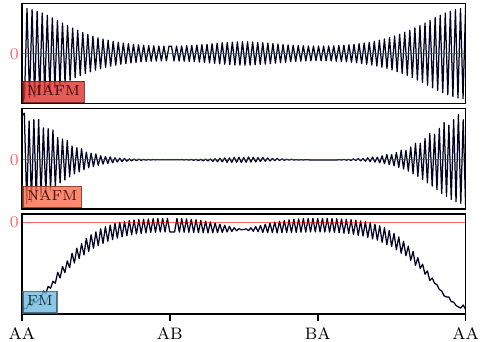}
\caption{Dominant magnetic orderings in twisted bilayer graphene near the magic angle. Shown is a linecut of the magnetic order parameter (spin density) along the diagonal of the rhombus-shaped moir\'e unit cell. The linecut is chosen to include the atoms that are closest to the actual line connecting one AA region with the next. Thus, at some point, there will always be a switch from an A sublattice site to yet another A sublattice site which produces a slip in the ordering. Top panel: \AA{}ngström scale anti-ferromagnetic with a modulation on the moir\'e scale (MAFM). Middle panel: \AA{}ngström scale anti-ferromagnetic with nodes in the AB and BA regions (NAFM). Bottom panel: mostly ferromagnetic (FM) order. }
\label{fig:Ordering}
\end{figure}

Next, we calculate the interacting spin susceptibility from Hartree$+U$ theory as function of doping at a wide range of twist angles near the magic angle (0.96$\degree$, 0.99$\degree$, 1.02$\degree$, 1.05$\degree$, 1.08$\degree$, 1.12$\degree$, 1.16$\degree$ and 1.20$\degree$). Figure~\ref{fig:CU} compares the critical Hubbard parameter $\Ucrit$ without Hartree interactions (left panel) and with Hartree interactions (right panel) as function of twist angle and doping. To assess if the system undergoes a phase transition, $\Ucrit$ must be compared with the actual value of $U$ for a carbon $p_z$-orbital. In graphene, Wehling \textit{et al.}~\cite{SECI} and Schuler \textit{et al.}~\cite{OHP} found that $U \approx 4$~eV. We expect that screening from tBLG does not significantly alter this value, as the flat bands mainly screen long-ranged interactions~\cite{PHD_2,CCRPA}. Therefore, we assume a doping and twist angle independent value of $U \approx 4$~eV in the following analysis.

Without Hartree interactions (left panel of Fig.~\ref{fig:CU}), magnetic instabilities are found at twist angles ranging from $0.99\degree$ to $1.12\degree$. At the magic angle ($\theta=1.05\degree$), instabilities occur at all integer doping levels except $\nu=2$. At twist angles smaller or larger than the magic angle, instabilities are observed for a smaller set of doping levels. In particular, for $\theta=0.99\degree$ and $\theta=1.12\degree$, they only occur at $\nu=-1$. In general, the critical Hubbard parameters are smaller for hole doped systems because the lower-energy flat bands are somewhat flatter than the higher-energy ones in tight-binding. For twist angles larger than 1.2$\degree$, we find $\Ucrit \approx 5.5\,$eV, which is similar to the value predicted for untwisted bilayer graphene~\cite{LK_CH}.

When Hartree interactions are included (right panel of Fig.~\ref{fig:CU}), a qualitatively different behaviour of $\Ucrit$ is observed near the magic angle. In particular, the lowest values of $\Ucrit$ are now found for electron-doped systems. Very close to the magic angle, $\Ucrit$ is lowest for $\nu=1$. At twist angles somewhat smaller or larger than the magic angle, the lowest value of $\Ucrit$ is at $\nu=2$ and at $\theta=0.96\degree$ or $1.16\degree$ the minimum is at $\nu=3$. These findings can be understood from the Hartree theory band structures, as seen in Fig.~\ref{fig:BS_H}, which show that the doping level which gives rise to the flattest bands depends on the twist angle: at the magic angle the flattest bands are found at $\nu=\pm 1$, while at $\theta=1.20\degree$ the higher-energy bands are extremely flat at $\nu=\pm 3$. Figure~\ref{fig:CU} also shows that magnetic instabilities occur over a larger twist angle range when long-ranged Hartree interactions are included. Specifically, the Hartree$+U$ approach predicts such instabilities for a twist-angle window from $\theta=0.96\degree$ to $\theta=1.16\degree$. This larger critical twist angle window is consistent with experimental findings: recent transport and tunnelling experiments reported correlated phases in a twist angle range from $1.0\degree$ to $1.2\degree$~\cite{Balents2020,cao2020nematicity}.

\begin{figure}[th]
\centering
\includegraphics[width=0.98\columnwidth]{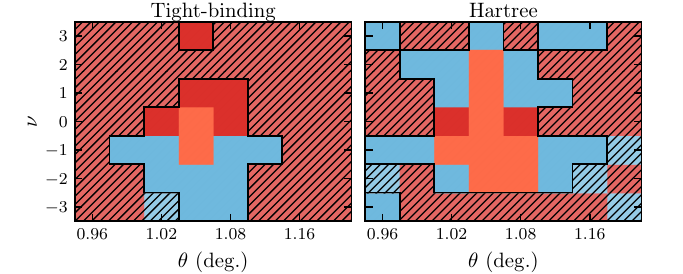}
\caption{Magnetic phase diagram of twisted bilayer graphene as function of flat band filling $\nu$ and twist angle $\theta$: blue denotes ferromagnetic order, while red and orange indicate modulated anti-ferromagnetic order and nodal anti-ferromagnetic order, respectively. Left panel: without Hartree interactions (tight-binding). Right panel: with Hartree interactions. Note that magnetic phases with $\Ucrit > 4\,\mathrm{eV}$ are experimentally not relevant (hatched regions).}
\label{fig:Zeta}
\end{figure}

Next, we analyze the spatial structure of the magnetic phases: the leading magnetic instabilities are either \AA{}ngström scale anti-ferromagnetic with a modulation on the moir\'e scale (MAFM), \AA{}ngström scale anti-ferromagnetic with nodes in the AB and BA regions (NAFM) or mostly ferromagnetic (FM), see Fig.~\ref{fig:Ordering}. Figure~\ref{fig:Zeta} shows the magnetic phase diagram as function of twist angle and doping near the magic angle. Without Hartree interactions, the hole doped system is typically FM. Ferromagnetism is found to coincide with small values of $\Ucrit$. In contrast, the undoped and electron doped system always exhibits MAFM, with NAFM only occurring at $\nu=0$ and $\nu=1$ at the magic angle.

Dramatic qualitative changes in the magnetic phase diagram are observed when Hartree interactions are included, see right panel of Fig.~\ref{fig:Ordering}. The region of NAFM order in $\nu-\theta$-space is larger, while MAFM is only found for the undoped system at $\theta=1.02\degree$ and $\theta=1.08\degree$. Everywhere else the ordering is FM. Again, occurrence of FM is correlated with low values of $\Ucrit$, which occur because of the interplay between the enhancement of the density of states from the long-ranged Hartree interactions upon doping and the enhancement of the density of states from changing the twist angle towards the magic angle.  

\section{Discussion}


In this section, we compare our calculated magnetic phase diagram to experimental findings. Many experimental techniques, including transport and tunnelling measurements, probe quasiparticle properties of tBLG. While our approach does not directly yield such properties, our analysis below reveals a strong correlation between the calculated value of the critical Hubbard parameter $\Ucrit$ and the measured quasiparticle gap in the correlated insulator phases, with small values of $\Ucrit$ corresponding to large gaps associated with pronounced resistive peaks in transport experiments. We stress that this correlation cannot be viewed as conclusive evidence that the experimentally observed correlated insulator states have a magnetic origin, because the large density of states at the Fermi which gives rise to the small values of $\Ucrit$ also promotes other instabilities (such as valley-ordered or nematic states).


At charge neutrality, our calculations predict small values of $\Ucrit$ near the magic angle with NAFM/MAFM order. Experimentally, the situation is not clear, however, with some experiments reporting semi-metallic behaviour near the magic angle~\cite{NAT_I,NAT_S,TSTBLG}, while others (for very similar twist angles) observe a strong insulating state~\cite{SOM}. These conflicting results could arise from different levels of strain in the samples: Liu \textit{et al.}~\cite{Liu2019} demonstrated that a $C_3$ broken symmetry state that is stabilized by strain retains its semi-metallic character because of the topological properties of the flat bands of tBLG.

Next, we consider the effect of doping. While at $\nu=-1$ insulating states are not often observed in experiments, some signatures of insulating states have been found at $\nu=+1$~\cite{TSTBLG,Vafek2020,Chern_Das,Polshyn2019}. This is consistent with our Hartree$+U$ results, which yield lower values of $\Ucrit$ for $\nu=+1$ than for $\nu=-1$. Note that the opposite result is obtained when long-ranged Hartree interactions are neglected. 

Experiments typically observe the strongest insulating states at $\nu=\pm2$~\cite{NAT_I,NAT_S,SOM,TSTBLG}. Without Hartree interactions, our calculations predict no broken-symmetry states at $\nu=+2$. In contrast, Hartree$+U$ theory predicts magnetic states for both $\nu=+2$ and $\nu=-2$. In recent experiments~\cite{Saito2020,Stepanov2020}, a thin dielectric spacer layer that separates the tBLG from metallic gates was used to enhance the screening of the electron-electron interactions in tBLG~\cite{PHD_3}. This results in significant changes to the electronic phase diagram with correlated insulator states being ``screened out" for most twist angles and doping levels~\cite{PHD_3}. Interestingly, these experiments often find the insulating state at $\nu=+2$ to be most robust. Naively, one might expect that this system should be described by the magnetic phase diagram obtained without long-ranged Hartree interactions. However, changes in external screening only result in small changes to the Hartree theory band structure~\cite{PHD_4,Cea2019} and therefore we expect that the Hartree$+U$ result should be more relevant to experiments with thin dielectric spacer layers. 


At $\nu=+3$, a strong insulating state is observed in experiments, especially when the tBLG is aligned with the hexagonal boron nitride substrate~\cite{EFM,Serlin2020}. In contrast, the $\nu=-3$ insulating state is almost never observed~\cite{SOM,TSTBLG,NAT_I,Chern_Das}. For insulating phases to emerge at these doping levels both valley and spin symmetries must be broken, i.e., the insulating state must be FM~\cite{EFM,Serlin2020}. This is consistent with the Hartree$+U$ results which predict FM order at $\nu=+3$ at several twist angles near the magic angle. Ferromagnetic order at $\nu=-3$ is only found at $\theta=0.96\degree$. Without Hartree interactions, our calculations do not predict FM order at $\nu=+3$ and instead we find relatively strong FM states at $\nu=-3$.

Hartree$+U$ theory also predicts that magnetic order at $\nu=+3$ should occur over a relatively large twist angle range, while those at $\nu=+1$ are only found very close to the magic angle. This finding also appears to be consistent with experiments. For example, Yankowitz \textit{et al.}~\cite{TSTBLG} observed an insulating state at $\nu=+3$ for a twist angle of $1.14\degree$, but no insulating state was found at $\nu=+1$. Interestingly, there are also clear signatures of this trend from recent scanning tunneling microscopy experiments of Choi~\textit{et al.}~\cite{Youngjoon2021}. At large twist angles, they observe that the $\nu=+3$ insulating state occurs before the $\nu=+1$ or $\nu=+2$. At slightly smaller twist angles, an additional insulating state at $\nu=+2$ occurs, with even smaller angles very close to the magic angle exhibiting insulating states for all integer electron doped systems. This observation is in very good agreement with our Hartree$+U$ results. Whereas, without Hartree interactions the opposite trend is observed: the leading instabilities occur closer to $\nu=-1$ for the largest angles away from the magic angle.

In summary, we observe a strong correlation between the critical values of the Hubbard interactions obtained from  Hartree$+U$ calculations and the experimentally measured quasiparticle gaps of the correlated insulator states. In contrast, no such correlation is observed when long-ranged Hartree interactions are neglected. 


As mentioned above, our current linear-response approach does not yield quasiparticle band structures of the broken-symmetry phases. In principle, such band structures can be obtained from self-consistent Hartree$+U$ calculations, but a qualitative picture can be derived from a symmetry analysis of the spatial structure of the leading magnetic instabilities. Importantly, neither the explicit mean-field calculations nor the symmetry analysis fully capture the effect of strong electron correlations on the quasiparticle band structure. For example, it is well known that strongly correlated electron systems can have energy gaps without any symmetry breaking (such gaps are induced by the frequency-dependence of the electron self-energy which is not captured by mean-field techniques). With this caveat in mind, we find that both MAFM and NAFM break the $C_{2}$ symmetry of tBLG, and therefore gap the flat band Dirac cone, which means NAFM and MAFM yield insulating states at charge neutrality~\cite{LK_CH}. Doping the MAFM and NAFM states with electrons or holes does not induce additional gaps and therefore the system is found to be metallic in agreement with explicit Hartree-Fock calculations~\cite{Cea2020}. The FM instability does not break $C_{2}$ (because the slight AFM character of the instability has a node between the AB and BA regions), but the spin degeneracy can be lifted and the bands can split to create an insulating state at charge neutrality. If the bands are spin split and doped away from charge neutrality, the system remains metallic as the $C_{2}$ symmetry is not broken. Therefore, this analysis only leads to insulating states at charge neutrality, while the doped magnetic states are found to be metallic. These results are in agreement with another atomistic calculation which found that only retaining Hubbard interactions can only yield insulating states at charge neutrality~\cite{Sboychakov2019,Sboychakov2020}, and also continuum model Hartree-Fock calculations that break $C_2$~\cite{Cea2020}. To overcome the limitations of the current approach, future research should investigate longer-ranged exchange interactions~\cite{SCHFC,Bultinck2020,Liu2019,Zhang2020,Cea2020,Stauber2020} and the influence of ordering tendencies with $\bvec q\neq0$ which could give rise to alternative symmetry breaking mechanisms such as valley~\cite{Wong2020} and rotational~\cite{NAT_CO} symmetry.

Finally, our Hartree$+U$ results for the magnetic phase diagram also have important implications for superconductivity in tBLG. First, band flattening induced by Hartree interactions enhances the density of states at the Fermi level and therefore increases the transition temperature irrespective of the nature of the superconducting glue. In addition, this mechanism also increases the range of twist angles where superconductivity can be observed~\cite{Lewandowski2021,Cea2021}. Note that superconductivity is typically observed in the vicinity of correlated insulator states at non-integer doping levels. Naively, one would expect that in this doping regime damped spin fluctuations from the magnetic parent state play an important role. However, Fischer and coworkers~\cite{AF2020} recently demonstrated the possibility of pairing by AFM spin fluctuations in the vicinity of a FM phase. Future work will investigate the predictions of Hartree$+U$ theory at non-integer doping levels to realise if long-ranged electron-electron interactions can also facilitate pairing by AFM spin fluctuations~\cite{Fischer_TTLG}.%

\section{Acknowledgements}

We are grateful for helpful discussions with V. Vitale, K. Atalar and Xia Liang. ZG was supported through a studentship in the Centre for Doctoral Training on Theory and Simulation of Materials at Imperial College London funded by the EPSRC (EP/L015579/1). We acknowledge funding from EPSRC grant EP/S025324/1 and the Thomas Young Centre under grant number TYC-101. We acknowledge the Imperial College London Research Computing Service (DOI:10.14469/hpc/2232) for the computational resources used in carrying out this work. The Deutsche  Forschungsgemeinschaft (DFG, German Research Foundation) is acknowledged for support through RTG 1995, within the Priority Program SPP 2244 “2DMP” and under Germany’s Excellence Strategy-Cluster of Excellence Matter and Light for Quantum Computing (ML4Q) EXC2004/1 - 390534769. We acknowledge support from the Max Planck-New York City Center for Non-Equilibrium Quantum Phenomena. Spin susceptibility calculations were performed with computing resources granted by RWTH Aachen University under projects rwth0496 and rwth0589.

\bibliographystyle{apsrev4-1}
\bibliography{REF}

\begin{thebibliography}{80}%
\makeatletter
\providecommand \@ifxundefined [1]{%
 \@ifx{#1\undefined}
}%
\providecommand \@ifnum [1]{%
 \ifnum #1\expandafter \@firstoftwo
 \else \expandafter \@secondoftwo
 \fi
}%
\providecommand \@ifx [1]{%
 \ifx #1\expandafter \@firstoftwo
 \else \expandafter \@secondoftwo
 \fi
}%
\providecommand \natexlab [1]{#1}%
\providecommand \enquote  [1]{``#1''}%
\providecommand \bibnamefont  [1]{#1}%
\providecommand \bibfnamefont [1]{#1}%
\providecommand \citenamefont [1]{#1}%
\providecommand \href@noop [0]{\@secondoftwo}%
\providecommand \href [0]{\begingroup \@sanitize@url \@href}%
\providecommand \@href[1]{\@@startlink{#1}\@@href}%
\providecommand \@@href[1]{\endgroup#1\@@endlink}%
\providecommand \@sanitize@url [0]{\catcode `\\12\catcode `\$12\catcode
  `\&12\catcode `\#12\catcode `\^12\catcode `\_12\catcode `\%12\relax}%
\providecommand \@@startlink[1]{}%
\providecommand \@@endlink[0]{}%
\providecommand \url  [0]{\begingroup\@sanitize@url \@url }%
\providecommand \@url [1]{\endgroup\@href {#1}{\urlprefix }}%
\providecommand \urlprefix  [0]{URL }%
\providecommand \Eprint [0]{\href }%
\providecommand \doibase [0]{http://dx.doi.org/}%
\providecommand \selectlanguage [0]{\@gobble}%
\providecommand \bibinfo  [0]{\@secondoftwo}%
\providecommand \bibfield  [0]{\@secondoftwo}%
\providecommand \translation [1]{[#1]}%
\providecommand \BibitemOpen [0]{}%
\providecommand \bibitemStop [0]{}%
\providecommand \bibitemNoStop [0]{.\EOS\space}%
\providecommand \EOS [0]{\spacefactor3000\relax}%
\providecommand \BibitemShut  [1]{\csname bibitem#1\endcsname}%
\let\auto@bib@innerbib\@empty
\bibitem [{\citenamefont {Cao}\ \emph {et~al.}(2018{\natexlab{a}})\citenamefont
  {Cao}, \citenamefont {Fatemi}, \citenamefont {Demir}, \citenamefont {Fang},
  \citenamefont {Tomarken}, \citenamefont {Luo}, \citenamefont
  {Sanchez-Yamagishi}, \citenamefont {Watanabe}, \citenamefont {Taniguchi},
  \citenamefont {Kaxiras}, \citenamefont {Ashoori},\ and\ \citenamefont
  {Jarillo-Herrero}}]{NAT_I}%
  \BibitemOpen
  \bibfield  {author} {\bibinfo {author} {\bibfnamefont {Y.}~\bibnamefont
  {Cao}}, \bibinfo {author} {\bibfnamefont {V.}~\bibnamefont {Fatemi}},
  \bibinfo {author} {\bibfnamefont {A.}~\bibnamefont {Demir}}, \bibinfo
  {author} {\bibfnamefont {S.}~\bibnamefont {Fang}}, \bibinfo {author}
  {\bibfnamefont {S.~L.}\ \bibnamefont {Tomarken}}, \bibinfo {author}
  {\bibfnamefont {J.~Y.}\ \bibnamefont {Luo}}, \bibinfo {author} {\bibfnamefont
  {J.~D.}\ \bibnamefont {Sanchez-Yamagishi}}, \bibinfo {author} {\bibfnamefont
  {K.}~\bibnamefont {Watanabe}}, \bibinfo {author} {\bibfnamefont
  {T.}~\bibnamefont {Taniguchi}}, \bibinfo {author} {\bibfnamefont
  {E.}~\bibnamefont {Kaxiras}}, \bibinfo {author} {\bibfnamefont {R.~C.}\
  \bibnamefont {Ashoori}}, \ and\ \bibinfo {author} {\bibfnamefont
  {P.}~\bibnamefont {Jarillo-Herrero}},\ }\href@noop {} {\bibfield  {journal}
  {\bibinfo  {journal} {Nature}\ }\textbf {\bibinfo {volume} {556}},\ \bibinfo
  {pages} {80} (\bibinfo {year} {2018}{\natexlab{a}})}\BibitemShut {NoStop}%
\bibitem [{\citenamefont {Cao}\ \emph {et~al.}(2018{\natexlab{b}})\citenamefont
  {Cao}, \citenamefont {Fatemi}, \citenamefont {Fang}, \citenamefont
  {Watanabe}, \citenamefont {Taniguchi}, \citenamefont {Kaxiras},\ and\
  \citenamefont {Jarillo-Herrero}}]{NAT_S}%
  \BibitemOpen
  \bibfield  {author} {\bibinfo {author} {\bibfnamefont {Y.}~\bibnamefont
  {Cao}}, \bibinfo {author} {\bibfnamefont {V.}~\bibnamefont {Fatemi}},
  \bibinfo {author} {\bibfnamefont {S.}~\bibnamefont {Fang}}, \bibinfo {author}
  {\bibfnamefont {K.}~\bibnamefont {Watanabe}}, \bibinfo {author}
  {\bibfnamefont {T.}~\bibnamefont {Taniguchi}}, \bibinfo {author}
  {\bibfnamefont {E.}~\bibnamefont {Kaxiras}}, \ and\ \bibinfo {author}
  {\bibfnamefont {P.}~\bibnamefont {Jarillo-Herrero}},\ }\href@noop {}
  {\bibfield  {journal} {\bibinfo  {journal} {Nature}\ }\textbf {\bibinfo
  {volume} {556}},\ \bibinfo {pages} {43} (\bibinfo {year}
  {2018}{\natexlab{b}})}\BibitemShut {NoStop}%
\bibitem [{\citenamefont {Carr}\ \emph {et~al.}(2017)\citenamefont {Carr},
  \citenamefont {Massatt}, \citenamefont {Fang}, \citenamefont {Cazeaux},
  \citenamefont {Luskin},\ and\ \citenamefont {Kaxiras}}]{TT}%
  \BibitemOpen
  \bibfield  {author} {\bibinfo {author} {\bibfnamefont {S.}~\bibnamefont
  {Carr}}, \bibinfo {author} {\bibfnamefont {D.}~\bibnamefont {Massatt}},
  \bibinfo {author} {\bibfnamefont {S.}~\bibnamefont {Fang}}, \bibinfo {author}
  {\bibfnamefont {P.}~\bibnamefont {Cazeaux}}, \bibinfo {author} {\bibfnamefont
  {M.}~\bibnamefont {Luskin}}, \ and\ \bibinfo {author} {\bibfnamefont
  {E.}~\bibnamefont {Kaxiras}},\ }\href@noop {} {\bibfield  {journal} {\bibinfo
   {journal} {Phys. Rev. B}\ }\textbf {\bibinfo {volume} {95}},\ \bibinfo
  {pages} {075420} (\bibinfo {year} {2017})}\BibitemShut {NoStop}%
\bibitem [{\citenamefont {Yankowitz}\ \emph {et~al.}(2019)\citenamefont
  {Yankowitz}, \citenamefont {Chen}, \citenamefont {Polshyn}, \citenamefont
  {Zhang}, \citenamefont {Watanabe}, \citenamefont {Taniguchi}, \citenamefont
  {Graf}, \citenamefont {Young},\ and\ \citenamefont {Dean}}]{TSTBLG}%
  \BibitemOpen
  \bibfield  {author} {\bibinfo {author} {\bibfnamefont {M.}~\bibnamefont
  {Yankowitz}}, \bibinfo {author} {\bibfnamefont {S.}~\bibnamefont {Chen}},
  \bibinfo {author} {\bibfnamefont {H.}~\bibnamefont {Polshyn}}, \bibinfo
  {author} {\bibfnamefont {Y.}~\bibnamefont {Zhang}}, \bibinfo {author}
  {\bibfnamefont {K.}~\bibnamefont {Watanabe}}, \bibinfo {author}
  {\bibfnamefont {T.}~\bibnamefont {Taniguchi}}, \bibinfo {author}
  {\bibfnamefont {D.}~\bibnamefont {Graf}}, \bibinfo {author} {\bibfnamefont
  {A.~F.}\ \bibnamefont {Young}}, \ and\ \bibinfo {author} {\bibfnamefont
  {C.~R.}\ \bibnamefont {Dean}},\ }\href@noop {} {\bibfield  {journal}
  {\bibinfo  {journal} {Science}\ }\textbf {\bibinfo {volume} {363}},\ \bibinfo
  {pages} {1059} (\bibinfo {year} {2019})}\BibitemShut {NoStop}%
\bibitem [{\citenamefont {Lu}\ \emph {et~al.}(2019)\citenamefont {Lu},
  \citenamefont {Stepanov}, \citenamefont {Yang}, \citenamefont {Xie},
  \citenamefont {Aamir}, \citenamefont {Das}, \citenamefont {Urgell},
  \citenamefont {Watanabe}, \citenamefont {Taniguchi}, \citenamefont {Zhang},
  \citenamefont {Bachtold}, \citenamefont {MacDonald},\ and\ \citenamefont
  {Efetov}}]{SOM}%
  \BibitemOpen
  \bibfield  {author} {\bibinfo {author} {\bibfnamefont {X.}~\bibnamefont
  {Lu}}, \bibinfo {author} {\bibfnamefont {P.}~\bibnamefont {Stepanov}},
  \bibinfo {author} {\bibfnamefont {W.}~\bibnamefont {Yang}}, \bibinfo {author}
  {\bibfnamefont {M.}~\bibnamefont {Xie}}, \bibinfo {author} {\bibfnamefont
  {M.~A.}\ \bibnamefont {Aamir}}, \bibinfo {author} {\bibfnamefont
  {I.}~\bibnamefont {Das}}, \bibinfo {author} {\bibfnamefont {C.}~\bibnamefont
  {Urgell}}, \bibinfo {author} {\bibfnamefont {K.}~\bibnamefont {Watanabe}},
  \bibinfo {author} {\bibfnamefont {T.}~\bibnamefont {Taniguchi}}, \bibinfo
  {author} {\bibfnamefont {G.}~\bibnamefont {Zhang}}, \bibinfo {author}
  {\bibfnamefont {A.}~\bibnamefont {Bachtold}}, \bibinfo {author}
  {\bibfnamefont {A.~H.}\ \bibnamefont {MacDonald}}, \ and\ \bibinfo {author}
  {\bibfnamefont {D.~K.}\ \bibnamefont {Efetov}},\ }\href@noop {} {\bibfield
  {journal} {\bibinfo  {journal} {Nature}\ }\textbf {\bibinfo {volume} {574}},\
  \bibinfo {pages} {653–657} (\bibinfo {year} {2019})}\BibitemShut {NoStop}%
\bibitem [{\citenamefont {Cao}\ \emph {et~al.}(2020{\natexlab{a}})\citenamefont
  {Cao}, \citenamefont {Rodan-Legrain}, \citenamefont {Park}, \citenamefont
  {Yuan}, \citenamefont {Watanabe}, \citenamefont {Taniguchi}, \citenamefont
  {Fernandes}, \citenamefont {Fu},\ and\ \citenamefont
  {Jarillo-Herrero}}]{cao2020nematicity}%
  \BibitemOpen
  \bibfield  {author} {\bibinfo {author} {\bibfnamefont {Y.}~\bibnamefont
  {Cao}}, \bibinfo {author} {\bibfnamefont {D.}~\bibnamefont {Rodan-Legrain}},
  \bibinfo {author} {\bibfnamefont {J.~M.}\ \bibnamefont {Park}}, \bibinfo
  {author} {\bibfnamefont {F.~N.}\ \bibnamefont {Yuan}}, \bibinfo {author}
  {\bibfnamefont {K.}~\bibnamefont {Watanabe}}, \bibinfo {author}
  {\bibfnamefont {T.}~\bibnamefont {Taniguchi}}, \bibinfo {author}
  {\bibfnamefont {R.~M.}\ \bibnamefont {Fernandes}}, \bibinfo {author}
  {\bibfnamefont {L.}~\bibnamefont {Fu}}, \ and\ \bibinfo {author}
  {\bibfnamefont {P.}~\bibnamefont {Jarillo-Herrero}},\ }\href@noop {}
  {\enquote {\bibinfo {title} {Nematicity and competing orders in
  superconducting magic-angle graphene},}\ } (\bibinfo {year}
  {2020}{\natexlab{a}}),\ \Eprint {http://arxiv.org/abs/2004.04148}
  {arXiv:2004.04148 [cond-mat.mes-hall]} \BibitemShut {NoStop}%
\bibitem [{\citenamefont {Zondiner}\ \emph {et~al.}(2020)\citenamefont
  {Zondiner}, \citenamefont {Rozen}, \citenamefont {Rodan-Legrain},
  \citenamefont {Cao}, \citenamefont {Queiroz}, \citenamefont {Taniguchi},
  \citenamefont {Watanabe}, \citenamefont {Oreg}, \citenamefont {von Oppen},
  \citenamefont {Stern}, \citenamefont {Berg}, \citenamefont
  {Jarillo-Herrero},\ and\ \citenamefont {Ilani}}]{Zondiner2020}%
  \BibitemOpen
  \bibfield  {author} {\bibinfo {author} {\bibfnamefont {U.}~\bibnamefont
  {Zondiner}}, \bibinfo {author} {\bibfnamefont {A.}~\bibnamefont {Rozen}},
  \bibinfo {author} {\bibfnamefont {D.}~\bibnamefont {Rodan-Legrain}}, \bibinfo
  {author} {\bibfnamefont {Y.}~\bibnamefont {Cao}}, \bibinfo {author}
  {\bibfnamefont {R.}~\bibnamefont {Queiroz}}, \bibinfo {author} {\bibfnamefont
  {T.}~\bibnamefont {Taniguchi}}, \bibinfo {author} {\bibfnamefont
  {K.}~\bibnamefont {Watanabe}}, \bibinfo {author} {\bibfnamefont
  {Y.}~\bibnamefont {Oreg}}, \bibinfo {author} {\bibfnamefont {F.}~\bibnamefont
  {von Oppen}}, \bibinfo {author} {\bibfnamefont {A.}~\bibnamefont {Stern}},
  \bibinfo {author} {\bibfnamefont {E.}~\bibnamefont {Berg}}, \bibinfo {author}
  {\bibfnamefont {P.}~\bibnamefont {Jarillo-Herrero}}, \ and\ \bibinfo {author}
  {\bibfnamefont {S.}~\bibnamefont {Ilani}},\ }\href@noop {} {\bibfield
  {journal} {\bibinfo  {journal} {Nature}\ }\textbf {\bibinfo {volume} {582}},\
  \bibinfo {pages} {203} (\bibinfo {year} {2020})}\BibitemShut {NoStop}%
\bibitem [{\citenamefont {Liu}\ \emph {et~al.}(2021{\natexlab{a}})\citenamefont
  {Liu}, \citenamefont {Wang}, \citenamefont {Watanabe}, \citenamefont
  {Taniguchi}, \citenamefont {Vafek},\ and\ \citenamefont {Li}}]{Vafek2020}%
  \BibitemOpen
  \bibfield  {author} {\bibinfo {author} {\bibfnamefont {X.}~\bibnamefont
  {Liu}}, \bibinfo {author} {\bibfnamefont {Z.}~\bibnamefont {Wang}}, \bibinfo
  {author} {\bibfnamefont {K.}~\bibnamefont {Watanabe}}, \bibinfo {author}
  {\bibfnamefont {T.}~\bibnamefont {Taniguchi}}, \bibinfo {author}
  {\bibfnamefont {O.}~\bibnamefont {Vafek}}, \ and\ \bibinfo {author}
  {\bibfnamefont {J.}~\bibnamefont {Li}},\ }\href@noop {} {\bibfield  {journal}
  {\bibinfo  {journal} {Science}\ }\textbf {\bibinfo {volume} {371}},\ \bibinfo
  {pages} {1261} (\bibinfo {year} {2021}{\natexlab{a}})}\BibitemShut {NoStop}%
\bibitem [{\citenamefont {Wong}\ \emph {et~al.}(2020)\citenamefont {Wong},
  \citenamefont {Nuckolls}, \citenamefont {Oh}, \citenamefont {Lian},
  \citenamefont {Yonglong~Xie}, \citenamefont {Watanabe}, \citenamefont
  {Taniguchi}, \citenamefont {Bernevig},\ and\ \citenamefont
  {Yazdani}}]{Wong2020}%
  \BibitemOpen
  \bibfield  {author} {\bibinfo {author} {\bibfnamefont {D.}~\bibnamefont
  {Wong}}, \bibinfo {author} {\bibfnamefont {K.~P.}\ \bibnamefont {Nuckolls}},
  \bibinfo {author} {\bibfnamefont {M.}~\bibnamefont {Oh}}, \bibinfo {author}
  {\bibfnamefont {B.}~\bibnamefont {Lian}}, \bibinfo {author} {\bibfnamefont
  {S.~J.}\ \bibnamefont {Yonglong~Xie}}, \bibinfo {author} {\bibfnamefont
  {K.}~\bibnamefont {Watanabe}}, \bibinfo {author} {\bibfnamefont
  {T.}~\bibnamefont {Taniguchi}}, \bibinfo {author} {\bibfnamefont {B.~A.}\
  \bibnamefont {Bernevig}}, \ and\ \bibinfo {author} {\bibfnamefont
  {A.}~\bibnamefont {Yazdani}},\ }\href@noop {} {\bibfield  {journal} {\bibinfo
   {journal} {Nature}\ }\textbf {\bibinfo {volume} {582}},\ \bibinfo {pages}
  {198–202} (\bibinfo {year} {2020})}\BibitemShut {NoStop}%
\bibitem [{\citenamefont {Xie}\ \emph {et~al.}(2019)\citenamefont {Xie},
  \citenamefont {Lian}, \citenamefont {J\"{a}ck}, \citenamefont {Liu},
  \citenamefont {Chiu}, \citenamefont {Watanabe}, \citenamefont {Taniguchi},
  \citenamefont {Bernevig},\ and\ \citenamefont {Yazdani}}]{NAT_SS}%
  \BibitemOpen
  \bibfield  {author} {\bibinfo {author} {\bibfnamefont {Y.}~\bibnamefont
  {Xie}}, \bibinfo {author} {\bibfnamefont {B.}~\bibnamefont {Lian}}, \bibinfo
  {author} {\bibfnamefont {B.}~\bibnamefont {J\"{a}ck}}, \bibinfo {author}
  {\bibfnamefont {X.}~\bibnamefont {Liu}}, \bibinfo {author} {\bibfnamefont
  {C.-L.}\ \bibnamefont {Chiu}}, \bibinfo {author} {\bibfnamefont
  {K.}~\bibnamefont {Watanabe}}, \bibinfo {author} {\bibfnamefont
  {T.}~\bibnamefont {Taniguchi}}, \bibinfo {author} {\bibfnamefont {B.~A.}\
  \bibnamefont {Bernevig}}, \ and\ \bibinfo {author} {\bibfnamefont
  {A.}~\bibnamefont {Yazdani}},\ }\href@noop {} {\bibfield  {journal} {\bibinfo
   {journal} {Nature}\ }\textbf {\bibinfo {volume} {572}},\ \bibinfo {pages}
  {101} (\bibinfo {year} {2019})}\BibitemShut {NoStop}%
\bibitem [{\citenamefont {Balents}\ \emph {et~al.}(2020)\citenamefont
  {Balents}, \citenamefont {Dean}, \citenamefont {Efetov},\ and\ \citenamefont
  {Young}}]{Balents2020}%
  \BibitemOpen
  \bibfield  {author} {\bibinfo {author} {\bibfnamefont {L.}~\bibnamefont
  {Balents}}, \bibinfo {author} {\bibfnamefont {C.~R.}\ \bibnamefont {Dean}},
  \bibinfo {author} {\bibfnamefont {D.~K.}\ \bibnamefont {Efetov}}, \ and\
  \bibinfo {author} {\bibfnamefont {A.~F.}\ \bibnamefont {Young}},\ }\href@noop
  {} {\bibfield  {journal} {\bibinfo  {journal} {Nat. Phys.}\ }\textbf
  {\bibinfo {volume} {16}},\ \bibinfo {pages} {725–733} (\bibinfo {year}
  {2020})}\BibitemShut {NoStop}%
\bibitem [{\citenamefont {Kennes}\ \emph {et~al.}(2021)\citenamefont {Kennes},
  \citenamefont {Claassen}, \citenamefont {Xian}, \citenamefont {Georges},
  \citenamefont {Millis}, \citenamefont {Hone}, \citenamefont {Dean},
  \citenamefont {Basov}, \citenamefont {Pasupathy},\ and\ \citenamefont
  {Rubio}}]{moiresim}%
  \BibitemOpen
  \bibfield  {author} {\bibinfo {author} {\bibfnamefont {D.~M.}\ \bibnamefont
  {Kennes}}, \bibinfo {author} {\bibfnamefont {M.}~\bibnamefont {Claassen}},
  \bibinfo {author} {\bibfnamefont {L.}~\bibnamefont {Xian}}, \bibinfo {author}
  {\bibfnamefont {A.}~\bibnamefont {Georges}}, \bibinfo {author} {\bibfnamefont
  {A.~J.}\ \bibnamefont {Millis}}, \bibinfo {author} {\bibfnamefont
  {J.}~\bibnamefont {Hone}}, \bibinfo {author} {\bibfnamefont {C.~R.}\
  \bibnamefont {Dean}}, \bibinfo {author} {\bibfnamefont {D.~N.}\ \bibnamefont
  {Basov}}, \bibinfo {author} {\bibfnamefont {A.}~\bibnamefont {Pasupathy}}, \
  and\ \bibinfo {author} {\bibfnamefont {A.}~\bibnamefont {Rubio}},\
  }\href@noop {} {\bibfield  {journal} {\bibinfo  {journal} {Nat. Phys.}\
  }\textbf {\bibinfo {volume} {17}},\ \bibinfo {pages} {155–163} (\bibinfo
  {year} {2021})}\BibitemShut {NoStop}%
\bibitem [{\citenamefont {Cao}\ \emph {et~al.}(2020{\natexlab{b}})\citenamefont
  {Cao}, \citenamefont {Chowdhury}, \citenamefont {Rodan-Legrain},
  \citenamefont {Rubies-Bigord\`a}, \citenamefont {Watanabe}, \citenamefont
  {Taniguchi}, \citenamefont {Senthil},\ and\ \citenamefont
  {Jarillo-Herrero}}]{SMTBLG}%
  \BibitemOpen
  \bibfield  {author} {\bibinfo {author} {\bibfnamefont {Y.}~\bibnamefont
  {Cao}}, \bibinfo {author} {\bibfnamefont {D.}~\bibnamefont {Chowdhury}},
  \bibinfo {author} {\bibfnamefont {D.}~\bibnamefont {Rodan-Legrain}}, \bibinfo
  {author} {\bibfnamefont {O.}~\bibnamefont {Rubies-Bigord\`a}}, \bibinfo
  {author} {\bibfnamefont {K.}~\bibnamefont {Watanabe}}, \bibinfo {author}
  {\bibfnamefont {T.}~\bibnamefont {Taniguchi}}, \bibinfo {author}
  {\bibfnamefont {T.}~\bibnamefont {Senthil}}, \ and\ \bibinfo {author}
  {\bibfnamefont {P.}~\bibnamefont {Jarillo-Herrero}},\ }\href@noop {}
  {\bibfield  {journal} {\bibinfo  {journal} {Phys. Rev. Lett.}\ }\textbf
  {\bibinfo {volume} {124}},\ \bibinfo {pages} {076801} (\bibinfo {year}
  {2020}{\natexlab{b}})}\BibitemShut {NoStop}%
\bibitem [{\citenamefont {Polshyn}\ \emph {et~al.}(2019)\citenamefont
  {Polshyn}, \citenamefont {Yankowitz}, \citenamefont {Chen}, \citenamefont
  {Zhang}, \citenamefont {Watanabe}, \citenamefont {Taniguchi}, \citenamefont
  {Dean},\ and\ \citenamefont {Young}}]{Polshyn2019}%
  \BibitemOpen
  \bibfield  {author} {\bibinfo {author} {\bibfnamefont {H.}~\bibnamefont
  {Polshyn}}, \bibinfo {author} {\bibfnamefont {M.}~\bibnamefont {Yankowitz}},
  \bibinfo {author} {\bibfnamefont {S.}~\bibnamefont {Chen}}, \bibinfo {author}
  {\bibfnamefont {Y.}~\bibnamefont {Zhang}}, \bibinfo {author} {\bibfnamefont
  {K.}~\bibnamefont {Watanabe}}, \bibinfo {author} {\bibfnamefont
  {T.}~\bibnamefont {Taniguchi}}, \bibinfo {author} {\bibfnamefont {C.~R.}\
  \bibnamefont {Dean}}, \ and\ \bibinfo {author} {\bibfnamefont {A.~F.}\
  \bibnamefont {Young}},\ }\href@noop {} {\bibfield  {journal} {\bibinfo
  {journal} {Nat. Phys.}\ }\textbf {\bibinfo {volume} {15}},\ \bibinfo {pages}
  {1011} (\bibinfo {year} {2019})}\BibitemShut {NoStop}%
\bibitem [{\citenamefont {Sharpe}\ \emph {et~al.}(2019)\citenamefont {Sharpe},
  \citenamefont {Fox}, \citenamefont {Barnard}, \citenamefont {Finney},
  \citenamefont {Watanabe}, \citenamefont {Taniguchi}, \citenamefont
  {Kastner},\ and\ \citenamefont {Goldhaber-Gordon}}]{EFM}%
  \BibitemOpen
  \bibfield  {author} {\bibinfo {author} {\bibfnamefont {A.~L.}\ \bibnamefont
  {Sharpe}}, \bibinfo {author} {\bibfnamefont {E.~J.}\ \bibnamefont {Fox}},
  \bibinfo {author} {\bibfnamefont {A.~W.}\ \bibnamefont {Barnard}}, \bibinfo
  {author} {\bibfnamefont {J.}~\bibnamefont {Finney}}, \bibinfo {author}
  {\bibfnamefont {K.}~\bibnamefont {Watanabe}}, \bibinfo {author}
  {\bibfnamefont {T.}~\bibnamefont {Taniguchi}}, \bibinfo {author}
  {\bibfnamefont {M.~A.}\ \bibnamefont {Kastner}}, \ and\ \bibinfo {author}
  {\bibfnamefont {D.}~\bibnamefont {Goldhaber-Gordon}},\ }\href@noop {}
  {\bibfield  {journal} {\bibinfo  {journal} {Science}\ }\textbf {\bibinfo
  {volume} {365}},\ \bibinfo {pages} {605–608} (\bibinfo {year}
  {2019})}\BibitemShut {NoStop}%
\bibitem [{\citenamefont {Serlin}\ \emph {et~al.}(2020)\citenamefont {Serlin},
  \citenamefont {Tschirhart}, \citenamefont {Polshyn}, \citenamefont {Zhang},
  \citenamefont {Zhu}, \citenamefont {Watanabe}, \citenamefont {Taniguchi},
  \citenamefont {Balents},\ and\ \citenamefont {Young}}]{Serlin2020}%
  \BibitemOpen
  \bibfield  {author} {\bibinfo {author} {\bibfnamefont {M.}~\bibnamefont
  {Serlin}}, \bibinfo {author} {\bibfnamefont {C.~L.}\ \bibnamefont
  {Tschirhart}}, \bibinfo {author} {\bibfnamefont {H.}~\bibnamefont {Polshyn}},
  \bibinfo {author} {\bibfnamefont {Y.}~\bibnamefont {Zhang}}, \bibinfo
  {author} {\bibfnamefont {J.}~\bibnamefont {Zhu}}, \bibinfo {author}
  {\bibfnamefont {K.}~\bibnamefont {Watanabe}}, \bibinfo {author}
  {\bibfnamefont {T.}~\bibnamefont {Taniguchi}}, \bibinfo {author}
  {\bibfnamefont {L.}~\bibnamefont {Balents}}, \ and\ \bibinfo {author}
  {\bibfnamefont {A.~F.}\ \bibnamefont {Young}},\ }\href@noop {} {\bibfield
  {journal} {\bibinfo  {journal} {Science}\ }\textbf {\bibinfo {volume}
  {367}},\ \bibinfo {pages} {900–903} (\bibinfo {year} {2020})}\BibitemShut
  {NoStop}%
\bibitem [{\citenamefont {Saito}\ \emph {et~al.}(2020)\citenamefont {Saito},
  \citenamefont {Ge}, \citenamefont {Watanabe}, \citenamefont {Taniguchi},\
  and\ \citenamefont {Young}}]{Saito2020}%
  \BibitemOpen
  \bibfield  {author} {\bibinfo {author} {\bibfnamefont {Y.}~\bibnamefont
  {Saito}}, \bibinfo {author} {\bibfnamefont {J.}~\bibnamefont {Ge}}, \bibinfo
  {author} {\bibfnamefont {K.}~\bibnamefont {Watanabe}}, \bibinfo {author}
  {\bibfnamefont {T.}~\bibnamefont {Taniguchi}}, \ and\ \bibinfo {author}
  {\bibfnamefont {A.~F.}\ \bibnamefont {Young}},\ }\href@noop {} {\bibfield
  {journal} {\bibinfo  {journal} {Nat. Phys.}\ }\textbf {\bibinfo {volume}
  {16}},\ \bibinfo {pages} {926} (\bibinfo {year} {2020})}\BibitemShut
  {NoStop}%
\bibitem [{\citenamefont {Arora}\ \emph {et~al.}(2020)\citenamefont {Arora},
  \citenamefont {Polski}, \citenamefont {Zhang}, \citenamefont {Thomson},
  \citenamefont {Choi}, \citenamefont {Kim}, \citenamefont {Lin}, \citenamefont
  {Wilson}, \citenamefont {Xu}, \citenamefont {Chu}, \citenamefont {Watanabe},
  \citenamefont {Taniguchi}, \citenamefont {Alicea},\ and\ \citenamefont
  {Nadj-Perge}}]{Arora2020}%
  \BibitemOpen
  \bibfield  {author} {\bibinfo {author} {\bibfnamefont {H.~S.}\ \bibnamefont
  {Arora}}, \bibinfo {author} {\bibfnamefont {R.}~\bibnamefont {Polski}},
  \bibinfo {author} {\bibfnamefont {Y.}~\bibnamefont {Zhang}}, \bibinfo
  {author} {\bibfnamefont {A.}~\bibnamefont {Thomson}}, \bibinfo {author}
  {\bibfnamefont {Y.}~\bibnamefont {Choi}}, \bibinfo {author} {\bibfnamefont
  {H.}~\bibnamefont {Kim}}, \bibinfo {author} {\bibfnamefont {Z.}~\bibnamefont
  {Lin}}, \bibinfo {author} {\bibfnamefont {I.~Z.}\ \bibnamefont {Wilson}},
  \bibinfo {author} {\bibfnamefont {X.}~\bibnamefont {Xu}}, \bibinfo {author}
  {\bibfnamefont {J.-H.}\ \bibnamefont {Chu}}, \bibinfo {author} {\bibfnamefont
  {K.}~\bibnamefont {Watanabe}}, \bibinfo {author} {\bibfnamefont
  {T.}~\bibnamefont {Taniguchi}}, \bibinfo {author} {\bibfnamefont
  {J.}~\bibnamefont {Alicea}}, \ and\ \bibinfo {author} {\bibfnamefont
  {S.}~\bibnamefont {Nadj-Perge}},\ }\href@noop {} {\bibfield  {journal}
  {\bibinfo  {journal} {Nature}\ }\textbf {\bibinfo {volume} {583}},\ \bibinfo
  {pages} {379–384} (\bibinfo {year} {2020})}\BibitemShut {NoStop}%
\bibitem [{\citenamefont {Stepanov}\ \emph {et~al.}(2020)\citenamefont
  {Stepanov}, \citenamefont {Das}, \citenamefont {Lu}, \citenamefont
  {Fahimniya}, \citenamefont {Watanabe}, \citenamefont {Taniguchi},
  \citenamefont {Koppens}, \citenamefont {Lischner}, \citenamefont {Levitov},\
  and\ \citenamefont {Efetov}}]{Stepanov2020}%
  \BibitemOpen
  \bibfield  {author} {\bibinfo {author} {\bibfnamefont {P.}~\bibnamefont
  {Stepanov}}, \bibinfo {author} {\bibfnamefont {I.}~\bibnamefont {Das}},
  \bibinfo {author} {\bibfnamefont {X.}~\bibnamefont {Lu}}, \bibinfo {author}
  {\bibfnamefont {A.}~\bibnamefont {Fahimniya}}, \bibinfo {author}
  {\bibfnamefont {K.}~\bibnamefont {Watanabe}}, \bibinfo {author}
  {\bibfnamefont {T.}~\bibnamefont {Taniguchi}}, \bibinfo {author}
  {\bibfnamefont {F.~H.~L.}\ \bibnamefont {Koppens}}, \bibinfo {author}
  {\bibfnamefont {J.}~\bibnamefont {Lischner}}, \bibinfo {author}
  {\bibfnamefont {L.}~\bibnamefont {Levitov}}, \ and\ \bibinfo {author}
  {\bibfnamefont {D.~K.}\ \bibnamefont {Efetov}},\ }\href@noop {} {\bibfield
  {journal} {\bibinfo  {journal} {Nature}\ }\textbf {\bibinfo {volume} {583}},\
  \bibinfo {pages} {375–378} (\bibinfo {year} {2020})}\BibitemShut {NoStop}%
\bibitem [{\citenamefont {Das}\ \emph {et~al.}(2020)\citenamefont {Das},
  \citenamefont {Lu}, \citenamefont {Herzog-Arbeitman}, \citenamefont {Song},
  \citenamefont {Watanabe}, \citenamefont {Taniguchi}, \citenamefont
  {Bernevig},\ and\ \citenamefont {Efetov}}]{Chern_Das}%
  \BibitemOpen
  \bibfield  {author} {\bibinfo {author} {\bibfnamefont {I.}~\bibnamefont
  {Das}}, \bibinfo {author} {\bibfnamefont {X.}~\bibnamefont {Lu}}, \bibinfo
  {author} {\bibfnamefont {J.}~\bibnamefont {Herzog-Arbeitman}}, \bibinfo
  {author} {\bibfnamefont {Z.-D.}\ \bibnamefont {Song}}, \bibinfo {author}
  {\bibfnamefont {K.}~\bibnamefont {Watanabe}}, \bibinfo {author}
  {\bibfnamefont {T.}~\bibnamefont {Taniguchi}}, \bibinfo {author}
  {\bibfnamefont {B.~A.}\ \bibnamefont {Bernevig}}, \ and\ \bibinfo {author}
  {\bibfnamefont {D.~K.}\ \bibnamefont {Efetov}},\ }\href@noop {} {\bibfield
  {journal} {\bibinfo  {journal} {arXiv:2007.13390}\ } (\bibinfo {year}
  {2020})}\BibitemShut {NoStop}%
\bibitem [{\citenamefont {Wu}\ \emph {et~al.}(2020)\citenamefont {Wu},
  \citenamefont {Zhang}, \citenamefont {Watanabe}, \citenamefont {Taniguchi},\
  and\ \citenamefont {Andrei}}]{Chern_Wu}%
  \BibitemOpen
  \bibfield  {author} {\bibinfo {author} {\bibfnamefont {S.}~\bibnamefont
  {Wu}}, \bibinfo {author} {\bibfnamefont {Z.}~\bibnamefont {Zhang}}, \bibinfo
  {author} {\bibfnamefont {K.}~\bibnamefont {Watanabe}}, \bibinfo {author}
  {\bibfnamefont {T.}~\bibnamefont {Taniguchi}}, \ and\ \bibinfo {author}
  {\bibfnamefont {E.~Y.}\ \bibnamefont {Andrei}},\ }\href@noop {} {\bibfield
  {journal} {\bibinfo  {journal} {arXiv:2007.03735}\ } (\bibinfo {year}
  {2020})}\BibitemShut {NoStop}%
\bibitem [{\citenamefont {Nuckolls}\ \emph {et~al.}(2020)\citenamefont
  {Nuckolls}, \citenamefont {Oh}, \citenamefont {Wong}, \citenamefont {Lian},
  \citenamefont {Watanabe}, \citenamefont {Taniguchi}, \citenamefont
  {Bernevig},\ and\ \citenamefont {Yazdani}}]{Chern_Nuckolls}%
  \BibitemOpen
  \bibfield  {author} {\bibinfo {author} {\bibfnamefont {K.~P.}\ \bibnamefont
  {Nuckolls}}, \bibinfo {author} {\bibfnamefont {M.}~\bibnamefont {Oh}},
  \bibinfo {author} {\bibfnamefont {D.}~\bibnamefont {Wong}}, \bibinfo {author}
  {\bibfnamefont {B.}~\bibnamefont {Lian}}, \bibinfo {author} {\bibfnamefont
  {K.}~\bibnamefont {Watanabe}}, \bibinfo {author} {\bibfnamefont
  {T.}~\bibnamefont {Taniguchi}}, \bibinfo {author} {\bibfnamefont {B.~A.}\
  \bibnamefont {Bernevig}}, \ and\ \bibinfo {author} {\bibfnamefont
  {A.}~\bibnamefont {Yazdani}},\ }\href@noop {} {\bibfield  {journal} {\bibinfo
   {journal} {arXiv:2007.03810}\ } (\bibinfo {year} {2020})}\BibitemShut
  {NoStop}%
\bibitem [{\citenamefont {Kerelsky}\ \emph {et~al.}(2019)\citenamefont
  {Kerelsky}, \citenamefont {McGilly}, \citenamefont {Kennes}, \citenamefont
  {Xian}, \citenamefont {Yankowitz}, \citenamefont {Chen}, \citenamefont
  {Watanabe}, \citenamefont {Taniguchi}, \citenamefont {Hone}, \citenamefont
  {Dean}, \citenamefont {Rubio},\ and\ \citenamefont {Pasupathy}}]{NAT_MEI}%
  \BibitemOpen
  \bibfield  {author} {\bibinfo {author} {\bibfnamefont {A.}~\bibnamefont
  {Kerelsky}}, \bibinfo {author} {\bibfnamefont {L.~J.}\ \bibnamefont
  {McGilly}}, \bibinfo {author} {\bibfnamefont {D.~M.}\ \bibnamefont {Kennes}},
  \bibinfo {author} {\bibfnamefont {L.}~\bibnamefont {Xian}}, \bibinfo {author}
  {\bibfnamefont {M.}~\bibnamefont {Yankowitz}}, \bibinfo {author}
  {\bibfnamefont {S.}~\bibnamefont {Chen}}, \bibinfo {author} {\bibfnamefont
  {K.}~\bibnamefont {Watanabe}}, \bibinfo {author} {\bibfnamefont
  {T.}~\bibnamefont {Taniguchi}}, \bibinfo {author} {\bibfnamefont
  {J.}~\bibnamefont {Hone}}, \bibinfo {author} {\bibfnamefont {C.}~\bibnamefont
  {Dean}}, \bibinfo {author} {\bibfnamefont {A.}~\bibnamefont {Rubio}}, \ and\
  \bibinfo {author} {\bibfnamefont {A.~N.}\ \bibnamefont {Pasupathy}},\
  }\href@noop {} {\bibfield  {journal} {\bibinfo  {journal} {Nature}\ }\textbf
  {\bibinfo {volume} {572}},\ \bibinfo {pages} {95} (\bibinfo {year}
  {2019})}\BibitemShut {NoStop}%
\bibitem [{\citenamefont {Jiang}\ \emph {et~al.}(2019)\citenamefont {Jiang},
  \citenamefont {Lai}, \citenamefont {Watanabe}, \citenamefont {Taniguchi},
  \citenamefont {Haule}, \citenamefont {Mao},\ and\ \citenamefont
  {Andrei}}]{NAT_CO}%
  \BibitemOpen
  \bibfield  {author} {\bibinfo {author} {\bibfnamefont {Y.}~\bibnamefont
  {Jiang}}, \bibinfo {author} {\bibfnamefont {X.}~\bibnamefont {Lai}}, \bibinfo
  {author} {\bibfnamefont {K.}~\bibnamefont {Watanabe}}, \bibinfo {author}
  {\bibfnamefont {T.}~\bibnamefont {Taniguchi}}, \bibinfo {author}
  {\bibfnamefont {K.}~\bibnamefont {Haule}}, \bibinfo {author} {\bibfnamefont
  {J.}~\bibnamefont {Mao}}, \ and\ \bibinfo {author} {\bibfnamefont {E.~Y.}\
  \bibnamefont {Andrei}},\ }\href@noop {} {\bibfield  {journal} {\bibinfo
  {journal} {Nature}\ }\textbf {\bibinfo {volume} {573}},\ \bibinfo {pages}
  {91} (\bibinfo {year} {2019})}\BibitemShut {NoStop}%
\bibitem [{\citenamefont {Choi}\ \emph {et~al.}(2019)\citenamefont {Choi},
  \citenamefont {Kemmer}, \citenamefont {Peng}, \citenamefont {Thomson},
  \citenamefont {Arora}, \citenamefont {Polski}, \citenamefont {Zhang},
  \citenamefont {Ren}, \citenamefont {Alicea}, \citenamefont {Refael},
  \citenamefont {von Oppen}, \citenamefont {Watanabe}, \citenamefont
  {Taniguchi},\ and\ \citenamefont {Nadj-Perge}}]{IEC}%
  \BibitemOpen
  \bibfield  {author} {\bibinfo {author} {\bibfnamefont {Y.}~\bibnamefont
  {Choi}}, \bibinfo {author} {\bibfnamefont {J.}~\bibnamefont {Kemmer}},
  \bibinfo {author} {\bibfnamefont {Y.}~\bibnamefont {Peng}}, \bibinfo {author}
  {\bibfnamefont {A.}~\bibnamefont {Thomson}}, \bibinfo {author} {\bibfnamefont
  {H.}~\bibnamefont {Arora}}, \bibinfo {author} {\bibfnamefont
  {R.}~\bibnamefont {Polski}}, \bibinfo {author} {\bibfnamefont
  {Y.}~\bibnamefont {Zhang}}, \bibinfo {author} {\bibfnamefont
  {H.}~\bibnamefont {Ren}}, \bibinfo {author} {\bibfnamefont {J.}~\bibnamefont
  {Alicea}}, \bibinfo {author} {\bibfnamefont {G.}~\bibnamefont {Refael}},
  \bibinfo {author} {\bibfnamefont {F.}~\bibnamefont {von Oppen}}, \bibinfo
  {author} {\bibfnamefont {K.}~\bibnamefont {Watanabe}}, \bibinfo {author}
  {\bibfnamefont {T.}~\bibnamefont {Taniguchi}}, \ and\ \bibinfo {author}
  {\bibfnamefont {S.}~\bibnamefont {Nadj-Perge}},\ }\href@noop {} {\bibfield
  {journal} {\bibinfo  {journal} {Nat. Phys.}\ }\textbf {\bibinfo {volume}
  {15}},\ \bibinfo {pages} {1174} (\bibinfo {year} {2019})}\BibitemShut
  {NoStop}%
\bibitem [{\citenamefont {dos Santos}\ \emph {et~al.}(2007)\citenamefont {dos
  Santos}, \citenamefont {Peres},\ and\ \citenamefont {Neto}}]{GBWT}%
  \BibitemOpen
  \bibfield  {author} {\bibinfo {author} {\bibfnamefont {J.~M. B.~L.}\
  \bibnamefont {dos Santos}}, \bibinfo {author} {\bibfnamefont {N.~M.~R.}\
  \bibnamefont {Peres}}, \ and\ \bibinfo {author} {\bibfnamefont {A.~H.~C.}\
  \bibnamefont {Neto}},\ }\href@noop {} {\bibfield  {journal} {\bibinfo
  {journal} {Phys. Rev. Lett.}\ }\textbf {\bibinfo {volume} {99}},\ \bibinfo
  {pages} {256802} (\bibinfo {year} {2007})}\BibitemShut {NoStop}%
\bibitem [{\citenamefont {Bistritzer}\ and\ \citenamefont
  {MacDonald}(2010)}]{MBTBLG}%
  \BibitemOpen
  \bibfield  {author} {\bibinfo {author} {\bibfnamefont {R.}~\bibnamefont
  {Bistritzer}}\ and\ \bibinfo {author} {\bibfnamefont {A.~H.}\ \bibnamefont
  {MacDonald}},\ }\href@noop {} {\bibfield  {journal} {\bibinfo  {journal}
  {PNAS}\ }\textbf {\bibinfo {volume} {108}},\ \bibinfo {pages} {12233}
  (\bibinfo {year} {2010})}\BibitemShut {NoStop}%
\bibitem [{\citenamefont {de~Laissardi\`ere}\ \emph {et~al.}(2010)\citenamefont
  {de~Laissardi\`ere}, \citenamefont {Mayou},\ and\ \citenamefont
  {Magaud}}]{LDE}%
  \BibitemOpen
  \bibfield  {author} {\bibinfo {author} {\bibfnamefont {G.~T.}\ \bibnamefont
  {de~Laissardi\`ere}}, \bibinfo {author} {\bibfnamefont {D.}~\bibnamefont
  {Mayou}}, \ and\ \bibinfo {author} {\bibfnamefont {L.}~\bibnamefont
  {Magaud}},\ }\href@noop {} {\bibfield  {journal} {\bibinfo  {journal} {Nano
  Lett.}\ }\textbf {\bibinfo {volume} {10}},\ \bibinfo {pages} {804} (\bibinfo
  {year} {2010})}\BibitemShut {NoStop}%
\bibitem [{\citenamefont {de~Laissardi\`ere}\ \emph {et~al.}(2012)\citenamefont
  {de~Laissardi\`ere}, \citenamefont {Mayou},\ and\ \citenamefont
  {Magaud}}]{NSCS}%
  \BibitemOpen
  \bibfield  {author} {\bibinfo {author} {\bibfnamefont {G.~T.}\ \bibnamefont
  {de~Laissardi\`ere}}, \bibinfo {author} {\bibfnamefont {D.}~\bibnamefont
  {Mayou}}, \ and\ \bibinfo {author} {\bibfnamefont {L.}~\bibnamefont
  {Magaud}},\ }\href@noop {} {\bibfield  {journal} {\bibinfo  {journal} {Phys.
  Rev. B}\ }\textbf {\bibinfo {volume} {86}},\ \bibinfo {pages} {125413}
  (\bibinfo {year} {2012})}\BibitemShut {NoStop}%
\bibitem [{\citenamefont {Cantele}\ \emph {et~al.}(2020)\citenamefont
  {Cantele}, \citenamefont {Alf\`e}, \citenamefont {Conte}, \citenamefont
  {Cataudella}, \citenamefont {Ninno},\ and\ \citenamefont
  {Lucignano}}]{Cantele2020}%
  \BibitemOpen
  \bibfield  {author} {\bibinfo {author} {\bibfnamefont {G.}~\bibnamefont
  {Cantele}}, \bibinfo {author} {\bibfnamefont {D.}~\bibnamefont {Alf\`e}},
  \bibinfo {author} {\bibfnamefont {F.}~\bibnamefont {Conte}}, \bibinfo
  {author} {\bibfnamefont {V.}~\bibnamefont {Cataudella}}, \bibinfo {author}
  {\bibfnamefont {D.}~\bibnamefont {Ninno}}, \ and\ \bibinfo {author}
  {\bibfnamefont {P.}~\bibnamefont {Lucignano}},\ }\href@noop {} {\bibfield
  {journal} {\bibinfo  {journal} {arXiv:2004.14323v1}\ } (\bibinfo {year}
  {2020})}\BibitemShut {NoStop}%
\bibitem [{\citenamefont {Jain}\ \emph {et~al.}(2017)\citenamefont {Jain},
  \citenamefont {Juri\u{c}i\'c},\ and\ \citenamefont {Barkema}}]{STBBG}%
  \BibitemOpen
  \bibfield  {author} {\bibinfo {author} {\bibfnamefont {S.~K.}\ \bibnamefont
  {Jain}}, \bibinfo {author} {\bibfnamefont {V.}~\bibnamefont {Juri\u{c}i\'c}},
  \ and\ \bibinfo {author} {\bibfnamefont {G.~T.}\ \bibnamefont {Barkema}},\
  }\href@noop {} {\bibfield  {journal} {\bibinfo  {journal} {2D Mater.}\
  }\textbf {\bibinfo {volume} {4}},\ \bibinfo {pages} {015018} (\bibinfo {year}
  {2017})}\BibitemShut {NoStop}%
\bibitem [{\citenamefont {Gargiulo}\ and\ \citenamefont
  {Yazyev}(2018)}]{SETLA}%
  \BibitemOpen
  \bibfield  {author} {\bibinfo {author} {\bibfnamefont {F.}~\bibnamefont
  {Gargiulo}}\ and\ \bibinfo {author} {\bibfnamefont {O.~V.}\ \bibnamefont
  {Yazyev}},\ }\href@noop {} {\bibfield  {journal} {\bibinfo  {journal} {2D
  Mater.}\ }\textbf {\bibinfo {volume} {5}},\ \bibinfo {pages} {015019}
  (\bibinfo {year} {2018})}\BibitemShut {NoStop}%
\bibitem [{\citenamefont {Guinea}\ and\ \citenamefont {Walet}(2019)}]{CMLD}%
  \BibitemOpen
  \bibfield  {author} {\bibinfo {author} {\bibfnamefont {F.}~\bibnamefont
  {Guinea}}\ and\ \bibinfo {author} {\bibfnamefont {N.~R.}\ \bibnamefont
  {Walet}},\ }\href@noop {} {\bibfield  {journal} {\bibinfo  {journal} {Phys.
  Rev. B}\ }\textbf {\bibinfo {volume} {99}},\ \bibinfo {pages} {205134}
  (\bibinfo {year} {2019})}\BibitemShut {NoStop}%
\bibitem [{\citenamefont {Carr}\ \emph
  {et~al.}(2019{\natexlab{a}})\citenamefont {Carr}, \citenamefont {Fang},
  \citenamefont {Zhu},\ and\ \citenamefont {Kaxiras}}]{KDP}%
  \BibitemOpen
  \bibfield  {author} {\bibinfo {author} {\bibfnamefont {S.}~\bibnamefont
  {Carr}}, \bibinfo {author} {\bibfnamefont {S.}~\bibnamefont {Fang}}, \bibinfo
  {author} {\bibfnamefont {Z.}~\bibnamefont {Zhu}}, \ and\ \bibinfo {author}
  {\bibfnamefont {E.}~\bibnamefont {Kaxiras}},\ }\href@noop {} {\bibfield
  {journal} {\bibinfo  {journal} {Phys. Rev. Research}\ }\textbf {\bibinfo
  {volume} {1}},\ \bibinfo {pages} {013001} (\bibinfo {year}
  {2019}{\natexlab{a}})}\BibitemShut {NoStop}%
\bibitem [{\citenamefont {Koshino}\ \emph {et~al.}(2018)\citenamefont
  {Koshino}, \citenamefont {Yuan}, \citenamefont {Koretsune}, \citenamefont
  {Ochi}, \citenamefont {Kuroki},\ and\ \citenamefont {Fu}}]{MLWO}%
  \BibitemOpen
  \bibfield  {author} {\bibinfo {author} {\bibfnamefont {M.}~\bibnamefont
  {Koshino}}, \bibinfo {author} {\bibfnamefont {N.~F.~Q.}\ \bibnamefont
  {Yuan}}, \bibinfo {author} {\bibfnamefont {T.}~\bibnamefont {Koretsune}},
  \bibinfo {author} {\bibfnamefont {M.}~\bibnamefont {Ochi}}, \bibinfo {author}
  {\bibfnamefont {K.}~\bibnamefont {Kuroki}}, \ and\ \bibinfo {author}
  {\bibfnamefont {L.}~\bibnamefont {Fu}},\ }\href@noop {} {\bibfield  {journal}
  {\bibinfo  {journal} {Phys. Rev. X}\ }\textbf {\bibinfo {volume} {8}},\
  \bibinfo {pages} {031087} (\bibinfo {year} {2018})}\BibitemShut {NoStop}%
\bibitem [{\citenamefont {Kang}\ and\ \citenamefont {Vafek}(2018)}]{SMLWF}%
  \BibitemOpen
  \bibfield  {author} {\bibinfo {author} {\bibfnamefont {J.}~\bibnamefont
  {Kang}}\ and\ \bibinfo {author} {\bibfnamefont {O.}~\bibnamefont {Vafek}},\
  }\href@noop {} {\bibfield  {journal} {\bibinfo  {journal} {Phys. Rev. X}\
  }\textbf {\bibinfo {volume} {8}},\ \bibinfo {pages} {031088} (\bibinfo {year}
  {2018})}\BibitemShut {NoStop}%
\bibitem [{\citenamefont {Goodwin}\ \emph
  {et~al.}(2019{\natexlab{a}})\citenamefont {Goodwin}, \citenamefont
  {Corsetti}, \citenamefont {Mostofi},\ and\ \citenamefont {Lischner}}]{PHD_1}%
  \BibitemOpen
  \bibfield  {author} {\bibinfo {author} {\bibfnamefont {Z.~A.~H.}\
  \bibnamefont {Goodwin}}, \bibinfo {author} {\bibfnamefont {F.}~\bibnamefont
  {Corsetti}}, \bibinfo {author} {\bibfnamefont {A.~A.}\ \bibnamefont
  {Mostofi}}, \ and\ \bibinfo {author} {\bibfnamefont {J.}~\bibnamefont
  {Lischner}},\ }\href@noop {} {\bibfield  {journal} {\bibinfo  {journal}
  {Phys. Rev. B}\ }\textbf {\bibinfo {volume} {100}},\ \bibinfo {pages}
  {121106(R)} (\bibinfo {year} {2019}{\natexlab{a}})}\BibitemShut {NoStop}%
\bibitem [{\citenamefont {Po}\ \emph {et~al.}(2019)\citenamefont {Po},
  \citenamefont {Zou}, \citenamefont {Senthil},\ and\ \citenamefont
  {Vishwanath}}]{FTBM}%
  \BibitemOpen
  \bibfield  {author} {\bibinfo {author} {\bibfnamefont {H.~C.}\ \bibnamefont
  {Po}}, \bibinfo {author} {\bibfnamefont {L.}~\bibnamefont {Zou}}, \bibinfo
  {author} {\bibfnamefont {T.}~\bibnamefont {Senthil}}, \ and\ \bibinfo
  {author} {\bibfnamefont {A.}~\bibnamefont {Vishwanath}},\ }\href@noop {}
  {\bibfield  {journal} {\bibinfo  {journal} {Phys. Rev. B}\ }\textbf {\bibinfo
  {volume} {99}},\ \bibinfo {pages} {195455} (\bibinfo {year}
  {2019})}\BibitemShut {NoStop}%
\bibitem [{\citenamefont {Carr}\ \emph
  {et~al.}(2019{\natexlab{b}})\citenamefont {Carr}, \citenamefont {Fang},
  \citenamefont {Po}, \citenamefont {Vishwanath},\ and\ \citenamefont
  {Kaxiras}}]{Carr2019wannier}%
  \BibitemOpen
  \bibfield  {author} {\bibinfo {author} {\bibfnamefont {S.}~\bibnamefont
  {Carr}}, \bibinfo {author} {\bibfnamefont {S.}~\bibnamefont {Fang}}, \bibinfo
  {author} {\bibfnamefont {H.~C.}\ \bibnamefont {Po}}, \bibinfo {author}
  {\bibfnamefont {A.}~\bibnamefont {Vishwanath}}, \ and\ \bibinfo {author}
  {\bibfnamefont {E.}~\bibnamefont {Kaxiras}},\ }\href@noop {} {\bibfield
  {journal} {\bibinfo  {journal} {Phys. Rev. Research}\ }\textbf {\bibinfo
  {volume} {1}},\ \bibinfo {pages} {033072} (\bibinfo {year}
  {2019}{\natexlab{b}})}\BibitemShut {NoStop}%
\bibitem [{\citenamefont {Kennes}\ \emph {et~al.}(2018)\citenamefont {Kennes},
  \citenamefont {Lischner},\ and\ \citenamefont {Karrasch}}]{SCDID}%
  \BibitemOpen
  \bibfield  {author} {\bibinfo {author} {\bibfnamefont {D.~M.}\ \bibnamefont
  {Kennes}}, \bibinfo {author} {\bibfnamefont {J.}~\bibnamefont {Lischner}}, \
  and\ \bibinfo {author} {\bibfnamefont {C.}~\bibnamefont {Karrasch}},\
  }\href@noop {} {\bibfield  {journal} {\bibinfo  {journal} {Phys. Rev. B}\
  }\textbf {\bibinfo {volume} {98}},\ \bibinfo {pages} {241407(R)} (\bibinfo
  {year} {2018})}\BibitemShut {NoStop}%
\bibitem [{\citenamefont {Klebl}\ \emph {et~al.}(2020)\citenamefont {Klebl},
  \citenamefont {Kennes},\ and\ \citenamefont {Honerkamp}}]{LK_DMK_CH}%
  \BibitemOpen
  \bibfield  {author} {\bibinfo {author} {\bibfnamefont {L.}~\bibnamefont
  {Klebl}}, \bibinfo {author} {\bibfnamefont {D.~M.}\ \bibnamefont {Kennes}}, \
  and\ \bibinfo {author} {\bibfnamefont {C.}~\bibnamefont {Honerkamp}},\
  }\href@noop {} {\bibfield  {journal} {\bibinfo  {journal} {Phys. Rev. B}\
  }\textbf {\bibinfo {volume} {102}},\ \bibinfo {pages} {085109} (\bibinfo
  {year} {2020})}\BibitemShut {NoStop}%
\bibitem [{\citenamefont {Gonzalez-Arraga}\ \emph {et~al.}(2017)\citenamefont
  {Gonzalez-Arraga}, \citenamefont {Lado}, \citenamefont {Guinea},\ and\
  \citenamefont {San-Jose}}]{ECM}%
  \BibitemOpen
  \bibfield  {author} {\bibinfo {author} {\bibfnamefont {L.~A.}\ \bibnamefont
  {Gonzalez-Arraga}}, \bibinfo {author} {\bibfnamefont {J.~L.}\ \bibnamefont
  {Lado}}, \bibinfo {author} {\bibfnamefont {F.}~\bibnamefont {Guinea}}, \ and\
  \bibinfo {author} {\bibfnamefont {P.}~\bibnamefont {San-Jose}},\ }\href@noop
  {} {\bibfield  {journal} {\bibinfo  {journal} {Phys. Rev. Lett.}\ }\textbf
  {\bibinfo {volume} {119}},\ \bibinfo {pages} {107201} (\bibinfo {year}
  {2017})}\BibitemShut {NoStop}%
\bibitem [{\citenamefont {Klebl}\ and\ \citenamefont
  {Honerkamp}(2019)}]{LK_CH}%
  \BibitemOpen
  \bibfield  {author} {\bibinfo {author} {\bibfnamefont {L.}~\bibnamefont
  {Klebl}}\ and\ \bibinfo {author} {\bibfnamefont {C.}~\bibnamefont
  {Honerkamp}},\ }\href {\doibase 10.1103/PhysRevB.100.155145} {\bibfield
  {journal} {\bibinfo  {journal} {Phys. Rev. B}\ }\textbf {\bibinfo {volume}
  {100}},\ \bibinfo {pages} {155145} (\bibinfo {year} {2019})}\BibitemShut
  {NoStop}%
\bibitem [{\citenamefont {Ramires}\ and\ \citenamefont
  {Lado}(2019)}]{Ramires2019}%
  \BibitemOpen
  \bibfield  {author} {\bibinfo {author} {\bibfnamefont {A.}~\bibnamefont
  {Ramires}}\ and\ \bibinfo {author} {\bibfnamefont {J.~L.}\ \bibnamefont
  {Lado}},\ }\href@noop {} {\bibfield  {journal} {\bibinfo  {journal} {Phys.
  Rev. B}\ }\textbf {\bibinfo {volume} {99}},\ \bibinfo {pages} {245118}
  (\bibinfo {year} {2019})}\BibitemShut {NoStop}%
\bibitem [{Note1()}]{Note1}%
  \BibitemOpen
  \bibinfo {note} {Note that Hubbard models using a basis of flat-band Wannier
  functions account for some long-ranged interactions because of the large size
  of the Wannier orbitals~\cite {MLWO,SMLWF,PHD_1}}\BibitemShut {NoStop}%
\bibitem [{\citenamefont {Guinea}\ and\ \citenamefont {Walet}(2018)}]{EE}%
  \BibitemOpen
  \bibfield  {author} {\bibinfo {author} {\bibfnamefont {F.}~\bibnamefont
  {Guinea}}\ and\ \bibinfo {author} {\bibfnamefont {N.~R.}\ \bibnamefont
  {Walet}},\ }\href@noop {} {\bibfield  {journal} {\bibinfo  {journal} {PNAS}\
  }\textbf {\bibinfo {volume} {115}},\ \bibinfo {pages} {13174–13179}
  (\bibinfo {year} {2018})}\BibitemShut {NoStop}%
\bibitem [{\citenamefont {Cea}\ \emph {et~al.}(2019)\citenamefont {Cea},
  \citenamefont {Walet},\ and\ \citenamefont {Guinea}}]{Cea2019}%
  \BibitemOpen
  \bibfield  {author} {\bibinfo {author} {\bibfnamefont {T.}~\bibnamefont
  {Cea}}, \bibinfo {author} {\bibfnamefont {N.~R.}\ \bibnamefont {Walet}}, \
  and\ \bibinfo {author} {\bibfnamefont {F.}~\bibnamefont {Guinea}},\
  }\href@noop {} {\bibfield  {journal} {\bibinfo  {journal} {Phys. Rev. B}\
  }\textbf {\bibinfo {volume} {100}},\ \bibinfo {pages} {205113} (\bibinfo
  {year} {2019})}\BibitemShut {NoStop}%
\bibitem [{\citenamefont {Rademaker}\ \emph {et~al.}(2019)\citenamefont
  {Rademaker}, \citenamefont {Abanin},\ and\ \citenamefont
  {Mellado}}]{Rademaker2019}%
  \BibitemOpen
  \bibfield  {author} {\bibinfo {author} {\bibfnamefont {L.}~\bibnamefont
  {Rademaker}}, \bibinfo {author} {\bibfnamefont {D.~A.}\ \bibnamefont
  {Abanin}}, \ and\ \bibinfo {author} {\bibfnamefont {P.}~\bibnamefont
  {Mellado}},\ }\href@noop {} {\bibfield  {journal} {\bibinfo  {journal} {Phys.
  Rev. B}\ }\textbf {\bibinfo {volume} {100}},\ \bibinfo {pages} {205114}
  (\bibinfo {year} {2019})}\BibitemShut {NoStop}%
\bibitem [{\citenamefont {Goodwin}\ \emph
  {et~al.}(2020{\natexlab{a}})\citenamefont {Goodwin}, \citenamefont {Vitale},
  \citenamefont {Liang}, \citenamefont {Mostofi},\ and\ \citenamefont
  {Lischner}}]{PHD_4}%
  \BibitemOpen
  \bibfield  {author} {\bibinfo {author} {\bibfnamefont {Z.~A.~H.}\
  \bibnamefont {Goodwin}}, \bibinfo {author} {\bibfnamefont {V.}~\bibnamefont
  {Vitale}}, \bibinfo {author} {\bibfnamefont {X.}~\bibnamefont {Liang}},
  \bibinfo {author} {\bibfnamefont {A.~A.}\ \bibnamefont {Mostofi}}, \ and\
  \bibinfo {author} {\bibfnamefont {J.}~\bibnamefont {Lischner}},\ }\href@noop
  {} {\bibfield  {journal} {\bibinfo  {journal} {Electron. Struct.}\ }\textbf
  {\bibinfo {volume} {2}},\ \bibinfo {pages} {034001} (\bibinfo {year}
  {2020}{\natexlab{a}})}\BibitemShut {NoStop}%
\bibitem [{\citenamefont {Calder\'on}\ and\ \citenamefont
  {Bascones}(2020)}]{Bascones2020}%
  \BibitemOpen
  \bibfield  {author} {\bibinfo {author} {\bibfnamefont {M.}~\bibnamefont
  {Calder\'on}}\ and\ \bibinfo {author} {\bibfnamefont {E.}~\bibnamefont
  {Bascones}},\ }\href@noop {} {\bibfield  {journal} {\bibinfo  {journal}
  {Phys. Rev. B}\ }\textbf {\bibinfo {volume} {102}},\ \bibinfo {pages}
  {155149} (\bibinfo {year} {2020})}\BibitemShut {NoStop}%
\bibitem [{\citenamefont {Xie}\ and\ \citenamefont {MacDonald}(2020)}]{SCHFC}%
  \BibitemOpen
  \bibfield  {author} {\bibinfo {author} {\bibfnamefont {M.}~\bibnamefont
  {Xie}}\ and\ \bibinfo {author} {\bibfnamefont {A.~H.}\ \bibnamefont
  {MacDonald}},\ }\href@noop {} {\bibfield  {journal} {\bibinfo  {journal}
  {Phys. Rev. Lett.}\ }\textbf {\bibinfo {volume} {124}},\ \bibinfo {pages}
  {097601} (\bibinfo {year} {2020})}\BibitemShut {NoStop}%
\bibitem [{\citenamefont {Bultinck}\ \emph {et~al.}(2020)\citenamefont
  {Bultinck}, \citenamefont {Khalaf}, \citenamefont {Liu}, \citenamefont
  {Chatterjee}, \citenamefont {Vishwanath},\ and\ \citenamefont
  {Zaletel}}]{Bultinck2020}%
  \BibitemOpen
  \bibfield  {author} {\bibinfo {author} {\bibfnamefont {N.}~\bibnamefont
  {Bultinck}}, \bibinfo {author} {\bibfnamefont {E.}~\bibnamefont {Khalaf}},
  \bibinfo {author} {\bibfnamefont {S.}~\bibnamefont {Liu}}, \bibinfo {author}
  {\bibfnamefont {S.}~\bibnamefont {Chatterjee}}, \bibinfo {author}
  {\bibfnamefont {A.}~\bibnamefont {Vishwanath}}, \ and\ \bibinfo {author}
  {\bibfnamefont {M.~P.}\ \bibnamefont {Zaletel}},\ }\href@noop {} {\bibfield
  {journal} {\bibinfo  {journal} {Phys. Rev. X}\ }\textbf {\bibinfo {volume}
  {10}},\ \bibinfo {pages} {031034} (\bibinfo {year} {2020})}\BibitemShut
  {NoStop}%
\bibitem [{\citenamefont {Liu}\ \emph {et~al.}(2021{\natexlab{b}})\citenamefont
  {Liu}, \citenamefont {Khalaf}, \citenamefont {Lee},\ and\ \citenamefont
  {Vishwanath}}]{Liu2019}%
  \BibitemOpen
  \bibfield  {author} {\bibinfo {author} {\bibfnamefont {S.}~\bibnamefont
  {Liu}}, \bibinfo {author} {\bibfnamefont {E.}~\bibnamefont {Khalaf}},
  \bibinfo {author} {\bibfnamefont {J.~Y.}\ \bibnamefont {Lee}}, \ and\
  \bibinfo {author} {\bibfnamefont {A.}~\bibnamefont {Vishwanath}},\
  }\href@noop {} {\bibfield  {journal} {\bibinfo  {journal} {Phys. Rev.
  Research}\ }\textbf {\bibinfo {volume} {103}},\ \bibinfo {pages} {013033}
  (\bibinfo {year} {2021}{\natexlab{b}})}\BibitemShut {NoStop}%
\bibitem [{\citenamefont {Zhang}\ \emph {et~al.}(2020)\citenamefont {Zhang},
  \citenamefont {Jiang}, \citenamefont {Wang},\ and\ \citenamefont
  {Zhang}}]{Zhang2020}%
  \BibitemOpen
  \bibfield  {author} {\bibinfo {author} {\bibfnamefont {Y.}~\bibnamefont
  {Zhang}}, \bibinfo {author} {\bibfnamefont {K.}~\bibnamefont {Jiang}},
  \bibinfo {author} {\bibfnamefont {Z.}~\bibnamefont {Wang}}, \ and\ \bibinfo
  {author} {\bibfnamefont {F.}~\bibnamefont {Zhang}},\ }\href@noop {}
  {\bibfield  {journal} {\bibinfo  {journal} {Phys. Rev. B}\ }\textbf {\bibinfo
  {volume} {102}},\ \bibinfo {pages} {035136} (\bibinfo {year}
  {2020})}\BibitemShut {NoStop}%
\bibitem [{\citenamefont {Cea}\ and\ \citenamefont {Guinea}(2020)}]{Cea2020}%
  \BibitemOpen
  \bibfield  {author} {\bibinfo {author} {\bibfnamefont {T.}~\bibnamefont
  {Cea}}\ and\ \bibinfo {author} {\bibfnamefont {F.}~\bibnamefont {Guinea}},\
  }\href@noop {} {\bibfield  {journal} {\bibinfo  {journal} {Phys. Rev. B}\
  }\textbf {\bibinfo {volume} {102}},\ \bibinfo {pages} {045107} (\bibinfo
  {year} {2020})}\BibitemShut {NoStop}%
\bibitem [{\citenamefont {Gonz\'alez}\ and\ \citenamefont
  {Stauber}(2020)}]{Stauber2020}%
  \BibitemOpen
  \bibfield  {author} {\bibinfo {author} {\bibfnamefont {J.}~\bibnamefont
  {Gonz\'alez}}\ and\ \bibinfo {author} {\bibfnamefont {T.}~\bibnamefont
  {Stauber}},\ }\href@noop {} {\bibfield  {journal} {\bibinfo  {journal} {Phys.
  Rev. B}\ }\textbf {\bibinfo {volume} {102}},\ \bibinfo {pages} {081118(R)}
  (\bibinfo {year} {2020})}\BibitemShut {NoStop}%
\bibitem [{\citenamefont {Sboychakov}\ \emph {et~al.}(2019)\citenamefont
  {Sboychakov}, \citenamefont {Rozhkov}, \citenamefont {Rakhmanov},\ and\
  \citenamefont {Nori}}]{Sboychakov2019}%
  \BibitemOpen
  \bibfield  {author} {\bibinfo {author} {\bibfnamefont {A.~O.}\ \bibnamefont
  {Sboychakov}}, \bibinfo {author} {\bibfnamefont {A.~V.}\ \bibnamefont
  {Rozhkov}}, \bibinfo {author} {\bibfnamefont {A.~L.}\ \bibnamefont
  {Rakhmanov}}, \ and\ \bibinfo {author} {\bibfnamefont {F.}~\bibnamefont
  {Nori}},\ }\href@noop {} {\bibfield  {journal} {\bibinfo  {journal} {Phys.
  Rev. B}\ }\textbf {\bibinfo {volume} {100}},\ \bibinfo {pages} {045111}
  (\bibinfo {year} {2019})}\BibitemShut {NoStop}%
\bibitem [{\citenamefont {Sboychakov}\ \emph {et~al.}(2020)\citenamefont
  {Sboychakov}, \citenamefont {Rozhkov}, \citenamefont {Rakhmanov},\ and\
  \citenamefont {Nori}}]{Sboychakov2020}%
  \BibitemOpen
  \bibfield  {author} {\bibinfo {author} {\bibfnamefont {A.~O.}\ \bibnamefont
  {Sboychakov}}, \bibinfo {author} {\bibfnamefont {A.~V.}\ \bibnamefont
  {Rozhkov}}, \bibinfo {author} {\bibfnamefont {A.~L.}\ \bibnamefont
  {Rakhmanov}}, \ and\ \bibinfo {author} {\bibfnamefont {F.}~\bibnamefont
  {Nori}},\ }\href@noop {} {\bibfield  {journal} {\bibinfo  {journal} {Phys.
  Rev. B}\ }\textbf {\bibinfo {volume} {102}},\ \bibinfo {pages} {155142}
  (\bibinfo {year} {2020})}\BibitemShut {NoStop}%
\bibitem [{\citenamefont {Liang}\ \emph {et~al.}(2020)\citenamefont {Liang},
  \citenamefont {Goodwin}, \citenamefont {Vitale}, \citenamefont {Corsetti},
  \citenamefont {Mostofi},\ and\ \citenamefont {Lischner}}]{PHD_5}%
  \BibitemOpen
  \bibfield  {author} {\bibinfo {author} {\bibfnamefont {X.}~\bibnamefont
  {Liang}}, \bibinfo {author} {\bibfnamefont {Z.~A.~H.}\ \bibnamefont
  {Goodwin}}, \bibinfo {author} {\bibfnamefont {V.}~\bibnamefont {Vitale}},
  \bibinfo {author} {\bibfnamefont {F.}~\bibnamefont {Corsetti}}, \bibinfo
  {author} {\bibfnamefont {A.~A.}\ \bibnamefont {Mostofi}}, \ and\ \bibinfo
  {author} {\bibfnamefont {J.}~\bibnamefont {Lischner}},\ }\href@noop {}
  {\bibfield  {journal} {\bibinfo  {journal} {Phys. Rev. B}\ }\textbf {\bibinfo
  {volume} {102}},\ \bibinfo {pages} {155146} (\bibinfo {year}
  {2020})}\BibitemShut {NoStop}%
\bibitem [{\citenamefont {O’Connor}\ \emph {et~al.}(2015)\citenamefont
  {O’Connor}, \citenamefont {Andzelm},\ and\ \citenamefont
  {Robbins}}]{AIREBO}%
  \BibitemOpen
  \bibfield  {author} {\bibinfo {author} {\bibfnamefont {T.~C.}\ \bibnamefont
  {O’Connor}}, \bibinfo {author} {\bibfnamefont {J.}~\bibnamefont {Andzelm}},
  \ and\ \bibinfo {author} {\bibfnamefont {M.~O.}\ \bibnamefont {Robbins}},\
  }\href@noop {} {\bibfield  {journal} {\bibinfo  {journal} {J. Chem. Phys.}\
  }\textbf {\bibinfo {volume} {142}},\ \bibinfo {pages} {024903} (\bibinfo
  {year} {2015})}\BibitemShut {NoStop}%
\bibitem [{\citenamefont {Kolmogorov}\ and\ \citenamefont {Crespi}(2005)}]{KC}%
  \BibitemOpen
  \bibfield  {author} {\bibinfo {author} {\bibfnamefont {A.~N.}\ \bibnamefont
  {Kolmogorov}}\ and\ \bibinfo {author} {\bibfnamefont {V.~H.}\ \bibnamefont
  {Crespi}},\ }\href@noop {} {\bibfield  {journal} {\bibinfo  {journal} {Phys.
  Rev. B}\ }\textbf {\bibinfo {volume} {71}},\ \bibinfo {pages} {235415}
  (\bibinfo {year} {2005})}\BibitemShut {NoStop}%
\bibitem [{\citenamefont {Plimpton}(1995)}]{LAMMPS}%
  \BibitemOpen
  \bibfield  {author} {\bibinfo {author} {\bibfnamefont {S.}~\bibnamefont
  {Plimpton}},\ }\href@noop {} {\bibfield  {journal} {\bibinfo  {journal} {J.
  Comp. Phys.}\ }\textbf {\bibinfo {volume} {117}},\ \bibinfo {pages} {1}
  (\bibinfo {year} {1995})}\BibitemShut {NoStop}%
\bibitem [{\citenamefont {Goodwin}\ \emph
  {et~al.}(2019{\natexlab{b}})\citenamefont {Goodwin}, \citenamefont
  {Corsetti}, \citenamefont {Mostofi},\ and\ \citenamefont {Lischner}}]{PHD_2}%
  \BibitemOpen
  \bibfield  {author} {\bibinfo {author} {\bibfnamefont {Z.~A.~H.}\
  \bibnamefont {Goodwin}}, \bibinfo {author} {\bibfnamefont {F.}~\bibnamefont
  {Corsetti}}, \bibinfo {author} {\bibfnamefont {A.~A.}\ \bibnamefont
  {Mostofi}}, \ and\ \bibinfo {author} {\bibfnamefont {J.}~\bibnamefont
  {Lischner}},\ }\href@noop {} {\bibfield  {journal} {\bibinfo  {journal}
  {Phys. Rev. B}\ }\textbf {\bibinfo {volume} {100}},\ \bibinfo {pages}
  {235424} (\bibinfo {year} {2019}{\natexlab{b}})}\BibitemShut {NoStop}%
\bibitem [{\citenamefont {Pizarro}\ \emph {et~al.}(2019)\citenamefont
  {Pizarro}, \citenamefont {Rosner}, \citenamefont {Thomale}, \citenamefont
  {Valent},\ and\ \citenamefont {Wehling}}]{CCRPA}%
  \BibitemOpen
  \bibfield  {author} {\bibinfo {author} {\bibfnamefont {J.~M.}\ \bibnamefont
  {Pizarro}}, \bibinfo {author} {\bibfnamefont {M.}~\bibnamefont {Rosner}},
  \bibinfo {author} {\bibfnamefont {R.}~\bibnamefont {Thomale}}, \bibinfo
  {author} {\bibfnamefont {R.}~\bibnamefont {Valent}}, \ and\ \bibinfo {author}
  {\bibfnamefont {T.~O.}\ \bibnamefont {Wehling}},\ }\href@noop {} {\bibfield
  {journal} {\bibinfo  {journal} {Phys. Rev. B}\ }\textbf {\bibinfo {volume}
  {100}},\ \bibinfo {pages} {161102(R)} (\bibinfo {year} {2019})}\BibitemShut
  {NoStop}%
\bibitem [{\citenamefont {Cea}\ and\ \citenamefont {Guinea}(2021)}]{Cea2021}%
  \BibitemOpen
  \bibfield  {author} {\bibinfo {author} {\bibfnamefont {T.}~\bibnamefont
  {Cea}}\ and\ \bibinfo {author} {\bibfnamefont {F.}~\bibnamefont {Guinea}},\
  }\href@noop {} {\bibfield  {journal} {\bibinfo  {journal} {arXiv:2103.01815}\
  } (\bibinfo {year} {2021})}\BibitemShut {NoStop}%
\bibitem [{\citenamefont {Scherer}\ \emph {et~al.}(2012)\citenamefont
  {Scherer}, \citenamefont {Uebelacker},\ and\ \citenamefont
  {Honerkamp}}]{Scherer2012}%
  \BibitemOpen
  \bibfield  {author} {\bibinfo {author} {\bibfnamefont {M.~M.}\ \bibnamefont
  {Scherer}}, \bibinfo {author} {\bibfnamefont {S.}~\bibnamefont {Uebelacker}},
  \ and\ \bibinfo {author} {\bibfnamefont {C.}~\bibnamefont {Honerkamp}},\
  }\href@noop {} {\bibfield  {journal} {\bibinfo  {journal} {Phys. Rev. B}\
  }\textbf {\bibinfo {volume} {85}},\ \bibinfo {pages} {235408} (\bibinfo
  {year} {2012})}\BibitemShut {NoStop}%
\bibitem [{\citenamefont {Lang}\ \emph {et~al.}(2012)\citenamefont {Lang},
  \citenamefont {Meng}, \citenamefont {Scherer}, \citenamefont {Uebelacker},
  \citenamefont {Assaad}, \citenamefont {Muramatsu}, \citenamefont
  {Honerkamp},\ and\ \citenamefont {Wessel}}]{Lang2012}%
  \BibitemOpen
  \bibfield  {author} {\bibinfo {author} {\bibfnamefont {T.~C.}\ \bibnamefont
  {Lang}}, \bibinfo {author} {\bibfnamefont {Z.~Y.}\ \bibnamefont {Meng}},
  \bibinfo {author} {\bibfnamefont {M.~M.}\ \bibnamefont {Scherer}}, \bibinfo
  {author} {\bibfnamefont {S.}~\bibnamefont {Uebelacker}}, \bibinfo {author}
  {\bibfnamefont {F.~F.}\ \bibnamefont {Assaad}}, \bibinfo {author}
  {\bibfnamefont {A.}~\bibnamefont {Muramatsu}}, \bibinfo {author}
  {\bibfnamefont {C.}~\bibnamefont {Honerkamp}}, \ and\ \bibinfo {author}
  {\bibfnamefont {S.}~\bibnamefont {Wessel}},\ }\href@noop {} {\bibfield
  {journal} {\bibinfo  {journal} {Phys. Rev. Lett.}\ }\textbf {\bibinfo
  {volume} {109}},\ \bibinfo {pages} {126402} (\bibinfo {year}
  {2012})}\BibitemShut {NoStop}%
\bibitem [{\citenamefont {Honerkamp}(2008)}]{Honerkamp2008}%
  \BibitemOpen
  \bibfield  {author} {\bibinfo {author} {\bibfnamefont {C.}~\bibnamefont
  {Honerkamp}},\ }\href@noop {} {\bibfield  {journal} {\bibinfo  {journal}
  {Phys. Rev. Lett.}\ }\textbf {\bibinfo {volume} {100}},\ \bibinfo {pages}
  {146404} (\bibinfo {year} {2008})}\BibitemShut {NoStop}%
\bibitem [{\citenamefont {Slater}\ and\ \citenamefont {Koster}(1954)}]{SK}%
  \BibitemOpen
  \bibfield  {author} {\bibinfo {author} {\bibfnamefont {J.~C.}\ \bibnamefont
  {Slater}}\ and\ \bibinfo {author} {\bibfnamefont {G.~F.}\ \bibnamefont
  {Koster}},\ }\href@noop {} {\bibfield  {journal} {\bibinfo  {journal} {Phys.
  Rev.}\ }\textbf {\bibinfo {volume} {94}},\ \bibinfo {pages} {1498} (\bibinfo
  {year} {1954})}\BibitemShut {NoStop}%
\bibitem [{\citenamefont {Moon}\ and\ \citenamefont
  {Koshino}(2012)}]{Moon2012}%
  \BibitemOpen
  \bibfield  {author} {\bibinfo {author} {\bibfnamefont {P.}~\bibnamefont
  {Moon}}\ and\ \bibinfo {author} {\bibfnamefont {M.}~\bibnamefont {Koshino}},\
  }\href {\doibase 10.1103/PhysRevB.85.195458} {\bibfield  {journal} {\bibinfo
  {journal} {Phys. Rev. B}\ }\textbf {\bibinfo {volume} {85}},\ \bibinfo
  {pages} {195458} (\bibinfo {year} {2012})}\BibitemShut {NoStop}%
\bibitem [{\citenamefont {Goodwin}\ \emph
  {et~al.}(2020{\natexlab{b}})\citenamefont {Goodwin}, \citenamefont {Vitale},
  \citenamefont {Corsetti}, \citenamefont {Efetov}, \citenamefont {Mostofi},\
  and\ \citenamefont {Lischner}}]{PHD_3}%
  \BibitemOpen
  \bibfield  {author} {\bibinfo {author} {\bibfnamefont {Z.~A.~H.}\
  \bibnamefont {Goodwin}}, \bibinfo {author} {\bibfnamefont {V.}~\bibnamefont
  {Vitale}}, \bibinfo {author} {\bibfnamefont {F.}~\bibnamefont {Corsetti}},
  \bibinfo {author} {\bibfnamefont {D.}~\bibnamefont {Efetov}}, \bibinfo
  {author} {\bibfnamefont {A.~A.}\ \bibnamefont {Mostofi}}, \ and\ \bibinfo
  {author} {\bibfnamefont {J.}~\bibnamefont {Lischner}},\ }\href@noop {}
  {\bibfield  {journal} {\bibinfo  {journal} {Phys. Rev. B}\ }\textbf {\bibinfo
  {volume} {101}},\ \bibinfo {pages} {165110} (\bibinfo {year}
  {2020}{\natexlab{b}})}\BibitemShut {NoStop}%
\bibitem [{\citenamefont {Lischner}\ \emph {et~al.}(2015)\citenamefont
  {Lischner}, \citenamefont {Bazhirov}, \citenamefont {MacDonald},
  \citenamefont {Cohen},\ and\ \citenamefont {Louie}}]{Lischner2015}%
  \BibitemOpen
  \bibfield  {author} {\bibinfo {author} {\bibfnamefont {J.}~\bibnamefont
  {Lischner}}, \bibinfo {author} {\bibfnamefont {T.}~\bibnamefont {Bazhirov}},
  \bibinfo {author} {\bibfnamefont {A.~H.}\ \bibnamefont {MacDonald}}, \bibinfo
  {author} {\bibfnamefont {M.~L.}\ \bibnamefont {Cohen}}, \ and\ \bibinfo
  {author} {\bibfnamefont {S.~G.}\ \bibnamefont {Louie}},\ }\href@noop {}
  {\bibfield  {journal} {\bibinfo  {journal} {Phys. Rev. B}\ }\textbf {\bibinfo
  {volume} {91}},\ \bibinfo {pages} {020502(R)} (\bibinfo {year}
  {2015})}\BibitemShut {NoStop}%
\bibitem [{Note2()}]{Note2}%
  \BibitemOpen
  \bibinfo {note} {The frequency grid is chosen to both be linearly spaced for
  $n \ll N_\omega $ ($n$ denotes the positive Matsubara frequency index ranging
  from zero to $N_\omega -1$) and increase its spacing quadratically for higher
  frequencies. This is achieved by $\omega _n\propto \protect \qopname \relax
  o{tan}[z_n\protect \tmspace +\thinmuskip {.1667em}\pi /2\protect \tmspace
  +\thinmuskip {.1667em}(2n+1)/(2N_\omega -1)]$ with $z_n=1-\epsilon _z
  [n/(N_\omega -1)]^{\alpha _z}$ controlling the strength of the divergence. In
  our case, we set $\epsilon _z = 10^{-7}$ and $\alpha _z=5$. The
  proportionality factor is chosen such that for small $n$, the original
  Matsubara frequencies are reproduced. The corresponding weights are
  determined by the derivative of the above formula with respect to
  $n$.}\BibitemShut {Stop}%
\bibitem [{\citenamefont {Lin}\ and\ \citenamefont
  {Hirsch}(1987)}]{LinHirschFM}%
  \BibitemOpen
  \bibfield  {author} {\bibinfo {author} {\bibfnamefont {H.~Q.}\ \bibnamefont
  {Lin}}\ and\ \bibinfo {author} {\bibfnamefont {J.~E.}\ \bibnamefont
  {Hirsch}},\ }\href {\doibase 10.1103/PhysRevB.35.3359} {\bibfield  {journal}
  {\bibinfo  {journal} {Phys. Rev. B}\ }\textbf {\bibinfo {volume} {35}},\
  \bibinfo {pages} {3359} (\bibinfo {year} {1987})}\BibitemShut {NoStop}%
\bibitem [{\citenamefont {Wehling}\ \emph {et~al.}(2011)\citenamefont
  {Wehling}, \citenamefont {\c{S}a\c{s}ıo\u{g}lu}, \citenamefont {Friedrich},
  \citenamefont {Lichtenstein}, \citenamefont {Katsnelson},\ and\ \citenamefont
  {Bl\"{u}gel}}]{SECI}%
  \BibitemOpen
  \bibfield  {author} {\bibinfo {author} {\bibfnamefont {T.~O.}\ \bibnamefont
  {Wehling}}, \bibinfo {author} {\bibfnamefont {E.}~\bibnamefont
  {\c{S}a\c{s}ıo\u{g}lu}}, \bibinfo {author} {\bibfnamefont {C.}~\bibnamefont
  {Friedrich}}, \bibinfo {author} {\bibfnamefont {A.~I.}\ \bibnamefont
  {Lichtenstein}}, \bibinfo {author} {\bibfnamefont {M.~I.}\ \bibnamefont
  {Katsnelson}}, \ and\ \bibinfo {author} {\bibfnamefont {S.}~\bibnamefont
  {Bl\"{u}gel}},\ }\href@noop {} {\bibfield  {journal} {\bibinfo  {journal}
  {Phys. Rev. Lett.}\ }\textbf {\bibinfo {volume} {106}},\ \bibinfo {pages}
  {236805} (\bibinfo {year} {2011})}\BibitemShut {NoStop}%
\bibitem [{\citenamefont {Sch\"{u}ler}\ \emph {et~al.}(2013)\citenamefont
  {Sch\"{u}ler}, \citenamefont {R\"{o}sner}, \citenamefont {Wehling},
  \citenamefont {Lichtenstein},\ and\ \citenamefont {Katsnelson}}]{OHP}%
  \BibitemOpen
  \bibfield  {author} {\bibinfo {author} {\bibfnamefont {M.}~\bibnamefont
  {Sch\"{u}ler}}, \bibinfo {author} {\bibfnamefont {M.}~\bibnamefont
  {R\"{o}sner}}, \bibinfo {author} {\bibfnamefont {T.~O.}\ \bibnamefont
  {Wehling}}, \bibinfo {author} {\bibfnamefont {A.~I.}\ \bibnamefont
  {Lichtenstein}}, \ and\ \bibinfo {author} {\bibfnamefont {M.~I.}\
  \bibnamefont {Katsnelson}},\ }\href@noop {} {\bibfield  {journal} {\bibinfo
  {journal} {Phys. Rev. Lett.}\ }\textbf {\bibinfo {volume} {111}},\ \bibinfo
  {pages} {036601} (\bibinfo {year} {2013})}\BibitemShut {NoStop}%
\bibitem [{\citenamefont {Choi}\ \emph {et~al.}(2021)\citenamefont {Choi},
  \citenamefont {Kim}, \citenamefont {Lewandowski}, \citenamefont {Peng},
  \citenamefont {Thomson}, \citenamefont {Polski}, \citenamefont {Zhang},
  \citenamefont {Watanabe}, \citenamefont {Taniguchi}, \citenamefont {Alicea},\
  and\ \citenamefont {Nadj-Perge}}]{Youngjoon2021}%
  \BibitemOpen
  \bibfield  {author} {\bibinfo {author} {\bibfnamefont {Y.}~\bibnamefont
  {Choi}}, \bibinfo {author} {\bibfnamefont {H.}~\bibnamefont {Kim}}, \bibinfo
  {author} {\bibfnamefont {C.}~\bibnamefont {Lewandowski}}, \bibinfo {author}
  {\bibfnamefont {Y.}~\bibnamefont {Peng}}, \bibinfo {author} {\bibfnamefont
  {A.}~\bibnamefont {Thomson}}, \bibinfo {author} {\bibfnamefont
  {R.}~\bibnamefont {Polski}}, \bibinfo {author} {\bibfnamefont
  {Y.}~\bibnamefont {Zhang}}, \bibinfo {author} {\bibfnamefont
  {K.}~\bibnamefont {Watanabe}}, \bibinfo {author} {\bibfnamefont
  {T.}~\bibnamefont {Taniguchi}}, \bibinfo {author} {\bibfnamefont
  {J.}~\bibnamefont {Alicea}}, \ and\ \bibinfo {author} {\bibfnamefont
  {S.}~\bibnamefont {Nadj-Perge}},\ }\href@noop {} {\bibfield  {journal}
  {\bibinfo  {journal} {arXiv:2102.02209}\ } (\bibinfo {year}
  {2021})}\BibitemShut {NoStop}%
\bibitem [{\citenamefont {Lewandowski}\ \emph {et~al.}(2021)\citenamefont
  {Lewandowski}, \citenamefont {Nadj-Perge},\ and\ \citenamefont
  {Chowdhury}}]{Lewandowski2021}%
  \BibitemOpen
  \bibfield  {author} {\bibinfo {author} {\bibfnamefont {C.}~\bibnamefont
  {Lewandowski}}, \bibinfo {author} {\bibfnamefont {S.}~\bibnamefont
  {Nadj-Perge}}, \ and\ \bibinfo {author} {\bibfnamefont {D.}~\bibnamefont
  {Chowdhury}},\ }\href@noop {} {\bibfield  {journal} {\bibinfo  {journal}
  {arXiv:2102.05661}\ } (\bibinfo {year} {2021})}\BibitemShut {NoStop}%
\bibitem [{\citenamefont {Fischer}\ \emph {et~al.}(2020)\citenamefont
  {Fischer}, \citenamefont {Klebl}, \citenamefont {Honerkamp},\ and\
  \citenamefont {Kennes}}]{AF2020}%
  \BibitemOpen
  \bibfield  {author} {\bibinfo {author} {\bibfnamefont {A.}~\bibnamefont
  {Fischer}}, \bibinfo {author} {\bibfnamefont {L.}~\bibnamefont {Klebl}},
  \bibinfo {author} {\bibfnamefont {C.}~\bibnamefont {Honerkamp}}, \ and\
  \bibinfo {author} {\bibfnamefont {D.~M.}\ \bibnamefont {Kennes}},\
  }\href@noop {} {\bibfield  {journal} {\bibinfo  {journal} {Phys. Rev. B}\
  }\textbf {\bibinfo {volume} {103}},\ \bibinfo {pages} {L041103} (\bibinfo
  {year} {2020})}\BibitemShut {NoStop}%
\bibitem [{\citenamefont {Fischer}\ \emph {et~al.}(2021)\citenamefont
  {Fischer}, \citenamefont {Goodwin}, \citenamefont {Mostofi}, \citenamefont
  {Lischner}, \citenamefont {Kennes},\ and\ \citenamefont
  {Klebl}}]{Fischer_TTLG}%
  \BibitemOpen
  \bibfield  {author} {\bibinfo {author} {\bibfnamefont {A.}~\bibnamefont
  {Fischer}}, \bibinfo {author} {\bibfnamefont {Z.~A.~H.}\ \bibnamefont
  {Goodwin}}, \bibinfo {author} {\bibfnamefont {A.~A.}\ \bibnamefont
  {Mostofi}}, \bibinfo {author} {\bibfnamefont {J.}~\bibnamefont {Lischner}},
  \bibinfo {author} {\bibfnamefont {D.~M.}\ \bibnamefont {Kennes}}, \ and\
  \bibinfo {author} {\bibfnamefont {L.}~\bibnamefont {Klebl}},\ }\href@noop {}
  {\bibfield  {journal} {\bibinfo  {journal} {arXiv:2104.10176}\ } (\bibinfo
  {year} {2021})}\BibitemShut {NoStop}%
\end{thebibliography}%

\end{document}